\documentclass[preprint,5p]{elsarticle}

\usepackage[T1]{fontenc}
\usepackage[utf8]{inputenc}
\usepackage{amsmath,amssymb,amsfonts}
\usepackage{svg}
\usepackage{graphicx}
\usepackage{textcomp}
\usepackage{xcolor}
\usepackage{caption}
\usepackage{pdfpages}
\usepackage{hyperref}
\usepackage{pythonhighlight}
\usepackage{courier}

\usepackage{listings}
\lstdefinestyle{C++}{
  language=C++,
  basicstyle=\ttfamily,
  breaklines=true,
  breakatwhitespace=true,
  commentstyle=\color{green},
}

\journal{JPDC}
\begin{document}

\begin{frontmatter}

\author[label1]{Dominik Ernst\corref{cor1}}
 \ead{dominik.ernst@fau.de}

\author[label2]{Markus Holzer}
 \author[label1]{Georg Hager}
 \author[label1]{Matthias Knorr}
 \author[label1]{Gerhard Wellein}

\affiliation[label1]{
  organization={Erlangen National High Performance Computing Center (NHR@FAU) \\ Friedrich-Alexander-Universität Erlangen-Nürnberg},
             addressline={\\Martensstraße 1},
             city={Erlangen},
            postcode={91058},
            country={Germany}
          }

\affiliation[label2]{organization = {Chair for System Simulation\\
Friedrich-Alexander-Universität Erlangen-Nürnberg},
country={Germany}
}

\title{Analytical Performance Estimation during Code Generation on Modern GPUs}

\begin{abstract}

Automatic code generation is frequently used to create
implementations of algorithms specifically tuned to particular
hardware and application parameters. The code generation process
involves the selection of adequate code transformations, tuning
parameters, and parallelization strategies. We propose an
alternative to time-intensive autotuning, scenario-specific
performance models, or black-box machine learning to select the
best-performing configuration.

This paper identifies the relevant performance-defining
mechanisms for memory-intensive GPU applications through a
performance model coupled with an analytic hardware
metric estimator. This enables a quick exploration of large
configuration spaces to identify highly efficient code candidates
with high accuracy.

We examine the changes of the A100 GPU architecture compared to
the predecessor V100 and address the challenges of how to model
the data transfer volumes through the new memory hierarchy.

We show how our method can be coupled to the ``pystencils''
stencil code generator, which is used to generate kernels for a
range-four 3D-25pt stencil and a complex two-phase fluid solver
based on the Lattice Boltzmann Method. For both, it delivers a
ranking that can be used to select the best-performing candidate.

The method is not limited to stencil kernels but can be
integrated into any code generator that can generate the required
address expressions.

\end{abstract}

\begin{highlights}
  \item Analytical Performance Modeling helps to find the best tuning parameters for code generation
  \item Analytical Performance Modeling gives insight that is unavailable to black box machine learning or auto tuning
  \item The thread block size is a decisive factor for GPU cache data volumes and performance
  \item GPU cache behavior and GPU performance can be accurately predicted based only on high level information
\end{highlights}


\end{frontmatter}

\section{Introduction}

\subsection{Spoilt for Choice in a Vast Configuration Space}

Code Generation is a programming technique that promises
increased developer productivity through more abstract program
representations and specialization to both the application scenario
and the hardware.

For the same algorithm specification, the code generation process
can make different decisions, leading to a large variety of
possible outcomes. These decisions could be different
parallelization variants, the fusion, ordering, and unrolling of loops, or
different blocking or tiling schemes, together with the associated
tuning parameters.  The influence of these decisions on the
performance is often not obvious, and therefore a common method to
find the best configuration is to use autotuning with a more or less
exhaustive search of the configuration space.

The evaluation of each configuration requires code generation and compilation
and benchmarking of the executable, which can take impractically long
for a large configuration space.

\begin{figure}
  \includegraphics[width=\columnwidth]{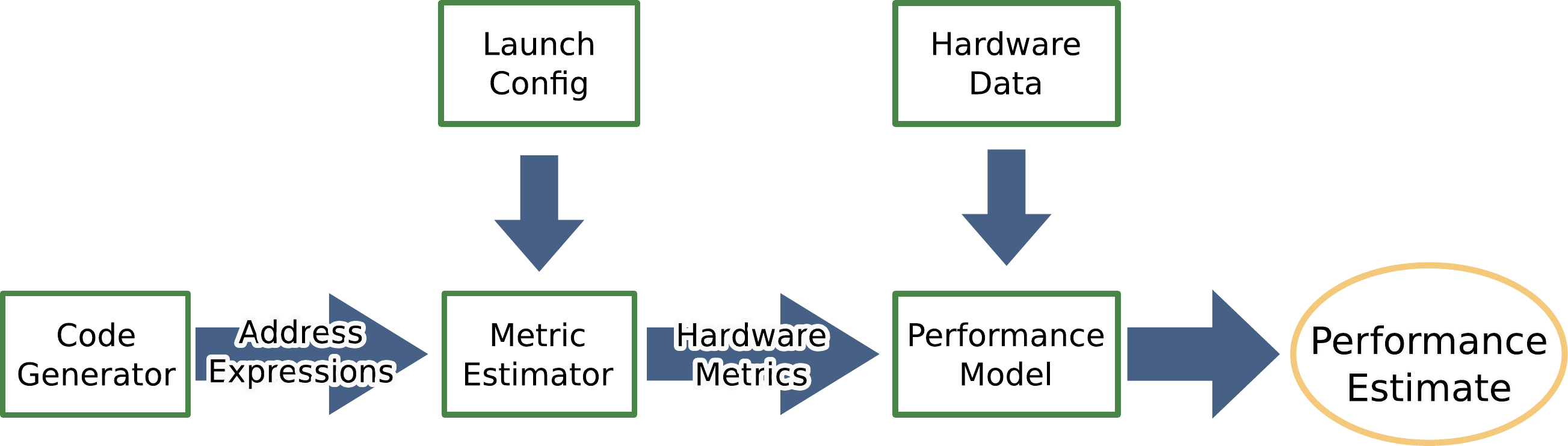}
  \caption{Illustration of the performance prediction workflow}
\label{fig:predictionflow}
\end{figure}

A way out of this dilemma is to stop treating generated code and its
performance as a black box but to make accurate
predictions instead about the resulting performance.
We present the metric and performance estimator \emph{Warpspeed}, which
couples metric estimation with a performance model, as illustrated in
figure \ref{fig:predictionflow}.

The main advantage of this approach is a much quicker evaluation time
compared to the generation-compilation-benchmark cycle.  We design the
hardware metric estimator with two goals in mind: It must only require
high-level features from the code generator that are already available
before the actual source code text is generated, and it must be
quickly evaluable.

Since there is no actual code execution in this process, it does not
require the target hardware, saving valuable resources especially
when high-end GPUs are involved. It also
enables the performance comparison of different GPU models, including
hypothetical GPUs for architectural exploration.
A performance model can not just evaluate different code configurations
on different hardware, but it can also grant insight into why the
performance is the way it is.  Knowing the performance-limiting
factors can inform further development of both algorithms and code
generation capabilities.

\subsection{Pystencils}

We show the utility of our performance estimator and how it
can be used in a code generator using the open-source python library \textit{pystencils} \cite{phasefields} as an example. With \textit{pystencils}, mathematical models can be directly described in an abstract representation. It is possible to automatically derive essential numerical schemes like the finite difference or the finite volume method from the abstract representation. Thus, complex mathematical models like multi-phase solidification models \cite{phasefields} can be described in a high-level latex-like representation close to usual descriptions in the literature. Furthermore, other packages like the Lattice-Boltzmann-Method (LBM) code generation framework lbmpy \cite{lbmpy1,lbmpy2} build on top of \textit{pystencils} to derive highly optimized numerical schemes for solving flow problems. From the numerical schemes, \textit{pystencils} then generates low-level C/OpenMP code to target CPUs or CUDA/OpenCL for GPUs.

As an example, consider the following 2D 4-point stencil:
\begin{python}
dst[0,0] = (src[1, 0] + src[-1, 0] +
            src[0, 1] + src[0, -1]) / 4
\end{python}

This example consists of a single assignment with relative field access to the destination and the source field dst and src.
\textit{Pystencils} then lowers this representation to its target dependent intermediate representation, where the relative field accesses are replaced with expressions computing the referenced addresses based on the thread coordinates, and the iteration space is realized either with explicit loops or mapped over hardware thread coordinates.
For example, for the access \pyth{src[1,0]} to the left neighbor, the referenced address would be computed on a GPU by this expression:

\begin{python}
  src_W = src +
     (threadIdx.x + blockIdx.x*blockDim.x + 1) +
     (threadIdx.y + blockIdx.y*blockDim.y) * w
\end{python}

Our performance estimator takes these address expressions, together with the launch configuration, field sizes and field alignments as the only high level information required from a code generator.
Our estimator is therefore not specific to \textit{pystencils}, but can be integrated with any code generation framework that is able to generate these artifacts.

The address expressions must contain only the base address of the field, and the thread and block coordinates as free fields.
These requirements on the address expressions also introduces limitations to the applicability of our performance estimator.
First, variables like grid sizes need to be known at generation time and indirect addressing is not supported.
It is possible to work around this using representative proxy values.

Second, control flow needs to be fully resolved beforehand. For very likely or unlikely, branches, it is often possible to simplify the control flow and still get a meaningful result.
The frequent \lstinline[style=C++]{if(tidx >= N)  return 0;} at the top of many GPU kernels, is false for almost all grid points and thus can be dropped without changing the result.

Another example are grid stride loops, where each thread would iterate over many points of the computational domain.
In this case, the code generator can emit the address expressions for just use one or a few iterations, and normalize the performance to the amount of work performed in these loop iterations.

\subsection{Contributions}

In this paper, we make the following significant contributions.  We
extend the roof{}line performance model for GPUs by two additional
limiters related to cache bandwidth.  We present a method to estimate
the cache hierarchy data volume metrics required for this
model based on the address expressions used to compute the memory
addresses referenced by a GPU program.  We propose the combination of
the model and the metric estimator for  use by code
generators, which can generate these address expressions and use the
combination to classify different code generation options and run
configurations.  In this context, we demonstrate the usefulness
of the method for distinguishing badly- from well-performing configurations and
identifying which type of configuration performs best on a A100 GPU on
two different, challenging applications implemented in the stencil
code generation framework \textit{pystencils}.

This work is an extended version of~\cite{SBAC}\@.  In
comparison to that paper we examine the Ampere GPU architecture and the changes introduced with the A100 GPU compared to the previously used V100 GPU and how its different cache architecture affects performance prediction.
We present a new method to estimate reuse from preceding threads in the
L2 cache, which became necessary due to the A100's larger L2 cache.
We reevaluate all measurements with the newer GPU.

We further demonstrate, that \emph{Layer Conditions} for codes with regular grid traversal can also be established for GPUs. Layer conditions are well know for CPUs where they are used, e.g., to estimate optimal blocking sizes for spatial and blocking of stencil codes.
To the best knowledge of the authors, similar effects have not yet been reported for GPU stencil calculations. The increased L2 cache capacity of the A100 GPU has made Layer Conditions more prominent than on previous GPUs with smaller L2 caches.

\subsection{Related Work}

There are numerous code generation frameworks for
stencils on GPUs.  These frameworks use either autotuning, or a performance model
that is specific to their requirements.
For example, the authors of \cite{libra} use a model to optimize the
computation/register ratio, which is important for the class of
stencils they are targeting.  In \cite{an5d},  a standard roof{}line model
with a fixed, theoretical memory volume is used for a full exploration of the configuration space, followed by benchmarking the top five candidates.

Optimization strategies like thread folding for stencils that have a similar range as ours are described in \cite{higherorderstencils} and \cite{higherorderstencils2}.

Similarly to our strategy, the framework LIFT \cite{lift1,lift2} extracts low-level features from an intermediate representation (IR).
Where our approach analytically derives a performance estimate from the extracted code features, they use a black box machine learning approach.
Machine learning approaches do not necessarily transfer well to scenarios not covered by the training data, and do not offer a clear view on underlying performance mechanisms.

The roof{}line model \cite{roofline_classic}, owes its popularity and wide
applicability to its simplicity, as it uses just two
performance limiters that apply to any architecture: 
memory bandwidth and peak floating-point performance.
These have been extended with cache-related limiters in the
\textit{hierarchical roof{}line model} \cite{roofline_hierarchical}
where data transfers volumes measured via hardware performance counters are
used to determine how close each memory level's bandwidth is to being a limiter.
NVIDIA's own \textit{nvvp} and \textit{nsight} profiling
applications do not just measure hardware quantities using performance
counters but also evaluate the measured results by comparing them
with the maximum possible rates.  This could be regarded as a multi-limiter
performance model that includes every limiter that is
measurable with performance counters.

In order to make a prediction, performance models focused on the memory hierarchy require application specific inputs in the form of the data volumes transferred in the memory hierarchy.

One work that attempts to estimate these metrics for GPUs is
\cite{reuse}. They put GPU simulator traces into a reuse-distance-based
model with time stepping that can also resolve timing-based effects, which
is potentially more precise in determining memory hierarchy data volumes than
our approach. They do however employ much more involved simulations and require
GPU execution traces.

The prediction accuracy can be improved further through architecture simulation. A detailed model of Volta's cache architecture is presented by \cite{Khairy_2020}. Though simulation is too slow for our usage scenario, it gives valuable insight into the mechanisms that have to be modeled.

The work presented in \cite{analyticalstencilmodel} targets the same usage
scenario and uses a similar performance model to ours. It also describes
capacity miss effects probabilistically and performs a similar evaluation
to ours.  The biggest difference is in using analytical formulas specific to a particular stencil to compute the memory volumes, which limits the applicability
to stencils, whereas our estimator does not have such a limitation.

Similarly to our approach, the work in \cite{warping} uses the \emph{Integer Set Library (ISL)} for symbolic set manipulation and counting. They perform detailed cache state simulation on CPUs based on information extracted from polyhedral models.

An auto tuning approach, which is an alternative to performance prediction as proposed by our method, is described by \cite{autotuning}. They show that an auto tuning approach can work without exhaustive scanning of the whole parameter space. Their tuning parameters include the thread block size and the thread folding configuration that we examine using our model.

The work in \cite{kernkraft} makes extensive use of the \emph{Layer Condition} for automatic, analytic metric and performance predictions on CPUs. The authors also published a website \cite{layerconditionexplorer}, containing a layer condition calculator, that allows to explore layer conditions in many scenarios.

\section{Performance Model}

The naive roof{}line model is often imprecise because
frequently, applications can saturate neither
of the two performance limits. In this work we
additionally consider the data transfers between L2 and L1
and between L1 and the registers, which may become the
bottlenecks instead in applications with effective cache reuse.

The four limiters considered here are  a reasonable selection out
of a wide range of options.
They yield a more differentiated performance ranking
by finding performance deviations between variants which
the roof{}line model would consider equivalent.
The selected set of limiters is general enough to be applicable
to all common GPU architectures, the
underlying these mechanisms can be reasoned about without too much
architectural inside knowledge, and the associated metrics can be
accurately estimated using high-level methods, a shown in this paper.

\section{Hardware Model}

The V100 data presented for comparison in this subsection is measured on a \verb'V100-PCIe-32GB' GPU, and the rest of the paper uses the \verb'A100-SXM4-40G' model, both with CUDA version $11.6$.

\begin{figure}
  \includegraphics[width=\columnwidth]{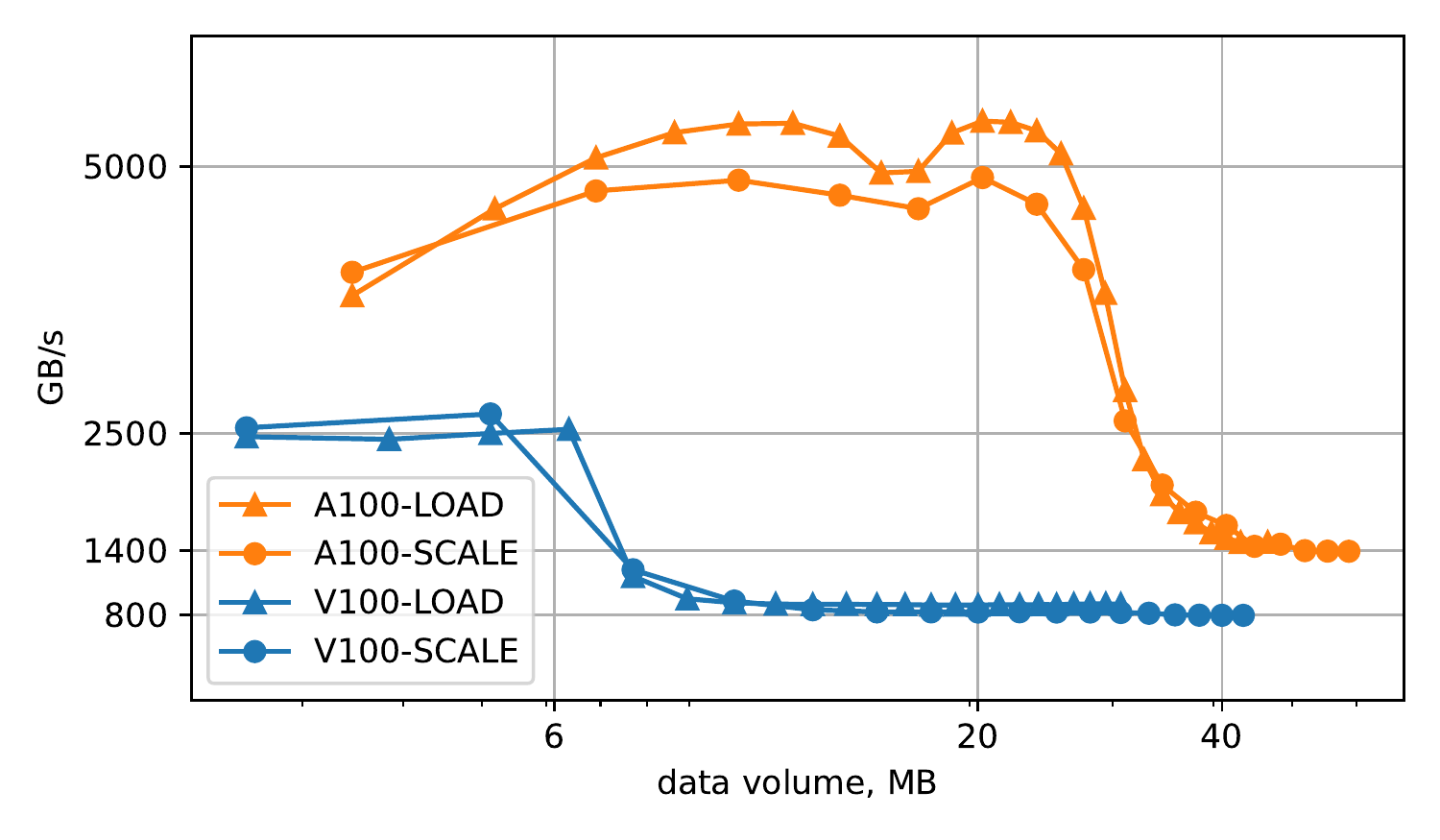}
  \caption{Data transfer rate vs data set size for two streaming kernels on a repeating data set. LOAD: \lstinline[style=C++] {c = A[i]}, SCALE: \lstinline[style=C++] {A[i] = c*B[i]}. Each thread's index \lstinline{i} is computed from the threads global thread index modulo the number of elements in the data set, resulting in repeated access to the same data by different threads.  }
  \label{fig:l2cachebw}
\end{figure}
\begin{figure}
  \includegraphics[width=\columnwidth]{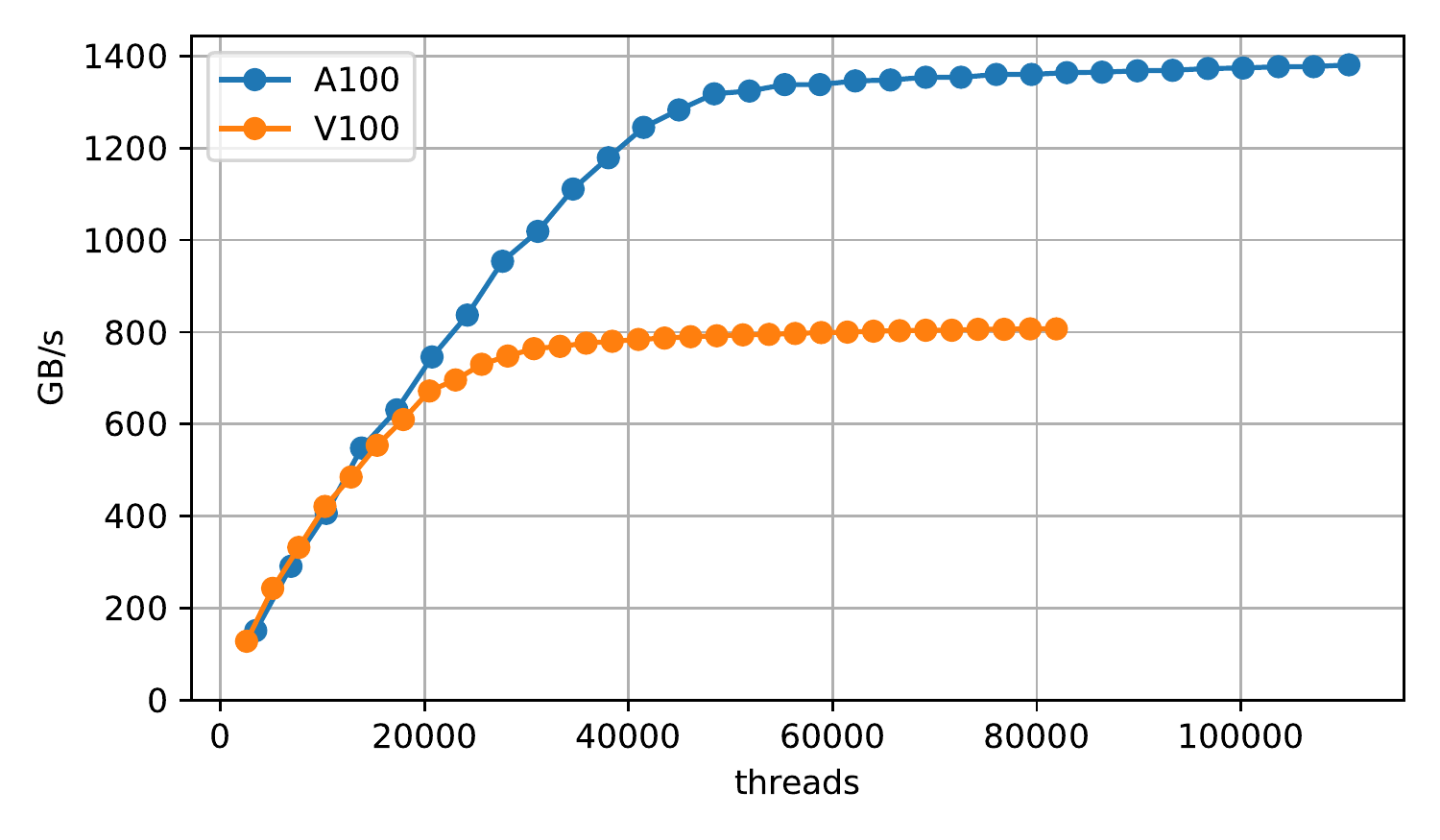}
  \caption{Data transfer rate vs thread count/occupancy for the SCALE kernel \lstinline[style=C++] {A[i] = c*B[i]}. The occupancy is artificially limited to two thread blocks per SM using a shared memory allocation as a spoiler. By adjusting the thread block size the occupancy is increased from $2\times32$ threads to full occupancy at $2\times1024$ threads per SM. }
  \label{fig:stream}
\end{figure}

The memory hierarchy found in NVIDIA's A100 GPU, featuring the Ampere architecture, has the same two level cache structure as its predecessor V100 with the Volta architecture, with a private L1 cache and a L2 cache that can be accessed by all \emph{Streaming Multiprocessors} (SM).
Table \ref{tab:av100} shows that the sizes of both caches, the number of SMs and the data transfer rates have changed, though not in equal proportions.

\begin{table}
  \begin{tabular}{lrrl}
                  & V100      & A100      &           \\ \hline
    SMs           & 80        & 108       & $+\, 35\%$   \\
     clocks, GHz   & 1.38  & 1.41   & $+\, 2\%$   \\
     L1 cache, kB     & 128    & 192     & $+\, 50\%$   \\
     L2 cache, MB     & 6      & $2\times20$     & $+\, 667\%$ \\
     DRAM BW, GB/s & 800       & 1400      & $+\, 75 \%$  \\
     L2 BW, GB/s        & 2500  & 5000 & $+\, 100\%$  \\
   \end{tabular}
   \caption{Table comparing the basic hardware properties of the \lstinline[style=C++]{V100-PCIe-32GB} and \lstinline[style=C++]{A100-SXM4-40G} GPUs}

   \label{tab:av100}
\end{table}

Though the 40MB L2 cache on the A100 is several times larger than on the V100, it is also split into two 20MB sections.
Each SM is connected to only one of the two L2 cache sections.
On a miss in the near L2 cache section, the far L2 cache section is checked, and the cache line is copied into the close L2 cache section in case of a hit.
This leads to duplicated L2 cache lines if data is accessed by two SMs belonging to two different cache sections.
In many scenarios, the effective L2 cache size is therefore halved to 20MB, which we will use as input for our model.

A link between the two L2 cache sections serves to maintain cache coherence.
While it would be possible to model and predict the amount of traffic over this link with our methods and use it as an additional L2 cache limiter, we have not observed many situations where that would be significant.
To exclude the data volume of this link in our comparisons of the accuracy of the predicted data volumes, we use the hardware performance counter \lstinline[style=C++]{lts__t_sectors_srcunit_tex_op_read} instead of the previously used counter \lstinline[style=C++]{lts__t_sectors_op_read}.
The qualifier \lstinline[style=C++]{srcunit_tex} only counts sectors requested by the SMs texture units, which is the quantity that our model estimates.

The measurement in figure \ref{fig:l2cachebw} shows data transfer rates for buffers of varying size repeatedly being accessed by different thread blocks.
The data shows that reliable L2 cache effectiveness can only be assumed for data sets less than 20MB.
The load-only and combined load-store kernel have similar transfer rates for both devices, which supports the assumption that there are no separate load/store paths.

Figure \ref{fig:stream} measures the data transfer rates of the DRAM for different amounts of concurrently running threads, or different amounts of occupancy.

Compared to the V100, the A100 increases the data transfer rates of the DRAM and the L2 cache by $75\%$ and $100\%$, whereas the computational, in-core resources, dependent on clock and SM count, increase by only $35\%$.
This altered balance is partly reflected in our performance model by a changed balance of the DRAM/L2 cache limiters compared to the L1 cache limiter.
However, this shift also increases the sensitivity of the performance to latency effects, which are not captured by our performance model.
Latency effects are the reason for the gradual ramp in performance in figure \ref{fig:stream}, where a higher occupancy is required to saturate the memory bandwidth.

In the larger L2 cache, data can be kept longer before being evicted.
This also extends the reuse opportunities from previously running threads further into the past.
As a consequence, the past threads reuse estimate that we described in \cite{SBAC} has to be extended to support the further look back.

\section{Metric Estimation}

\subsection{Floating Point Execution}

The number of floating-point operations, needed for the FP
throughput limiter, is straightforward to get from the source code.
The code generator performs common optimizations such as
constant folding and common subexpression elimination to anticipate
the changes that the compiler probably would do.
For the applications and configurations covered here, the floating-point
execution rate is never the limiter and will thus not be considered
for the rest of this paper.

\subsection{L1 Cache Load Throughput}

\begin{figure}
  \centering
  \includegraphics[width=\columnwidth]{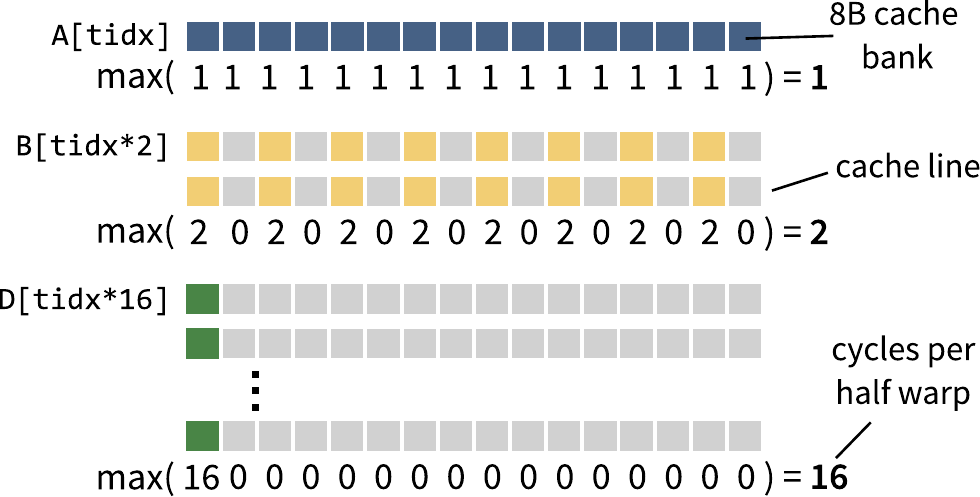}
  \caption{Illustration of cache bank conflicts. Each column is a cache bank, the numbers below are the number of addresses for that cache bank.}
  \label{fig:cachebanks}
\end{figure}

The time required to transfer data between L1 cache and registers is highly dependent on the access patterns, which is why it is not useful to just compute a transferred data volume and divide it by the maximum data transfer rate of the L1 cache.
Instead, we determine the number of cycles required to process each load/store instruction.
In the best case, the L1 cache in Volta and Ampere can access 128B per cycle, or one unique 8B double precision (DP) value for each thread in a 16-wide half warp.
That means that a single load/store instruction can be processed in as low as one cycle per half warp or two cycles for the full warp.
The 128B cache lines are distributed over 16 cache banks, each of which can deliver 8B per cycle.
A half warp memory access instruction can access data from different cache lines in a single cycle, as long as at most one value per cache bank is required.
Each additional address per cache bank requires an additional cycle.
A set of addresses that can be processed in the same cycle is called a wavefront.
Consider the four loads in the following listing and the illustration
in figure \ref{fig:cachebanks}:
\begin{lstlisting}[style=C++]
  double a = A[threadIdx.x];
  double b = B[threadIdx.x*2];
  double d = D[threadIdx.x*16];
\end{lstlisting}
The consecutive addresses in the load from A have no bank conflicts and result in one cycle per half warp.
Load B loads only every other 8B address, resulting in two addresses per cache bank, which are loaded in two wavefronts or cycles per half warp.
The worst case is represented by the load from D, where all addresses are in the same cache bank and have to be loaded over 16 cycles per half warp.

Additionally, we have found that two non-conflicting addresses that are too far apart cannot be processed in the same wavefront.
We haven't found an exact value for the distance of two addresses beyond which they cannot pair in a wavefront, only that ``close'' addresses generally do, and ``far'' addresses generally don't.
We use a distance of 1024B to discern between close and far apart addresses.

\begin{figure}
\begin{python}
def gridIteration(fields, threadGroup, visitor):
  for field in fields:
    addresses = []
    for addressExpr in field.accesses:
      for threadIdx in threadGroup:
        addresses.extend(addressExpr(threadIdx))
    visitor.count(addresses)

class BankConflictVisitor:
  cycles = 0
  def count(laneAddresses):
    banks = [0] * 16  # 16 banks
    # count references to each cache bank
    for a in unique(addresses):
      banks[(a // 8) 
    # bank with most references
    cycles += max(banks)
\end{python}
\caption {Simplified representation of the grid iteration function. The two example visitors compute the number of unique accessed 32B cache lines and the number of bank conflicts in the set of computed addresses.}
\label{lst:griditeration}
\end{figure}

To compute the total number of cycles that is being spent accessing data in the L1 cache, we use a generic grid iteration function as in listing \ref{lst:griditeration}.
This grid iteration function enumerates all thread indices of a group of threads, and computes all the addresses that result from putting these thread indices into all the address expressions of a field.

To compute the number of cycles required to process all the load/store instructions, we pass a visitor function, similar to the \pyth{BankConflictVisitor} function in the sample, to the grid iteration to customize it for that purpose.
This visitor computes the number of bank conflicts and therefore the number of cycles required to load the half warp's set of addresses computed by the grid iteration.
Averaging the results for all the half warps in a thread block makes the results more robust and less sensitive to edge cases.

The vectorized evaluation of the thread indices inside the thread groups using numpy's \pyth{meshgrid} function allows for fast enumeration of thread indices, so that the limiter for these enumerations is usually the \pyth{unique} function which is used to deduplicate addresses.

\subsection{L2 to L1 Cache Transfers}

\begin{figure}
  \centering
  \includegraphics[width=\columnwidth]{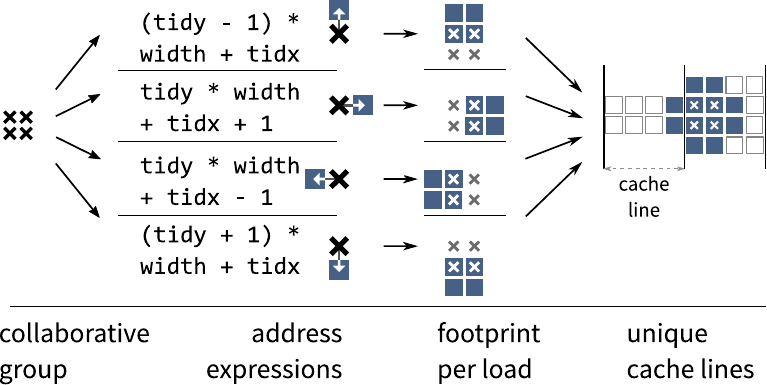}
  \caption{Illustration of the memory footprint computation. Example for a 2D 4 point stencil update and a $2\times2$ thread block.}
  \label{fig:footprint}
\end{figure}

The data volume that is transferred between the L1 and the L2 cache is determined by the effectiveness of the L1 cache.

A key property of memory accesses is their spatial and temporal locality.
For GPUs, temporal and spatial locality of a single thread's accesses is less relevant than the collaborative reuse among multiple threads sharing the L1 cache.
It is therefore not enough to look at a single thread in isolation to determine the amount of transferred data, but at the group of threads that share a cache level.
If the same address is referenced repeatedly by different threads that can share data in the L1 cache, the data at this address has to be loaded from the L2 cache only once.
The classification of cache misses by \cite{hennessy-patterson} calls these cache misses  \emph{compulsory}, because they are impossible to avoid.
This compulsory data volume can be computed by counting the number of unique cache lines that are referenced by a group of threads, which we call their \emph{unique memory footprint}.
It is necessary to count unique cache lines instead of addresses, because data is transferred at the granularity of cache lines.

The unique memory footprint of the four threads in Figure \ref{fig:footprint} after executing the four load operations, comprises six unique cache lines (white boxes), which contain the data of the 10 unique addresses (blue boxes) that are referenced by the 16 load operations.

\begin{figure}
  \centering
  \includegraphics[width=\columnwidth]{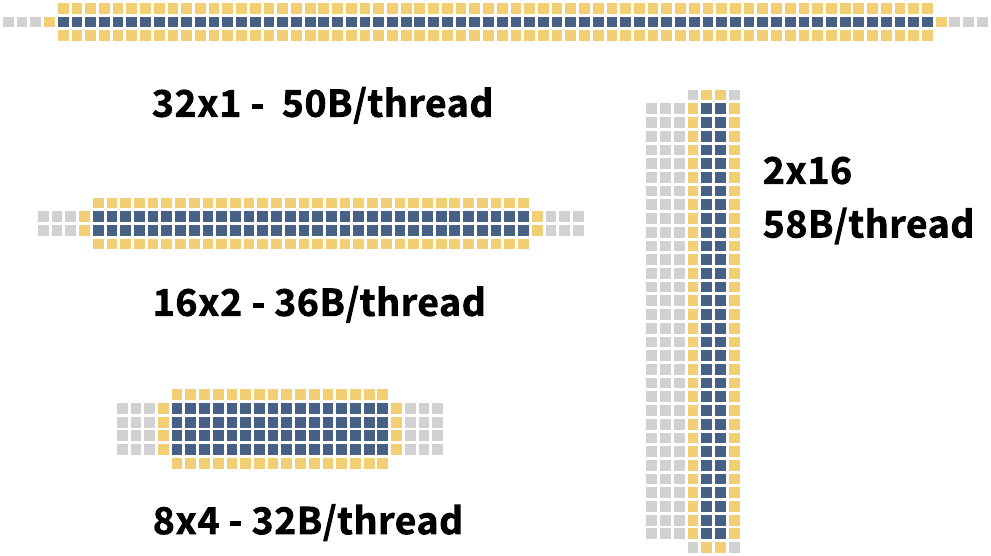}
  \caption{Illustration of how different thread block sizes result in different cache footprints for a 2D 5-point stencil. blue/yellow: internal/external point, gray: wasted cache line. Exact amount of wasted cache line allocation depends on the exact alignment of the thread block. }
  \label{fig:footprint_examples}
\end{figure}

Figure \ref{fig:footprint_examples} shows an illustration how different thread block sizes result in different memory footprints and data volumes per thread in the case of a 2D 5-point stencil.
Elongated shapes generally have higher data volumes than more quadratic ones due to their higher surface/volume ratio.
Moreover, cache line granularity in x direction favors wide shapes more than tall shapes.

Because threads in the same thread block can share data in the L1 cache, we compute the unique L1 cache footprint of a complete thread block as the data volume that this block transfers from the L2 cache.
The L2-L1 data volume per-thread is then the volume of the unique footprint divided by the number of threads in the block.

While threads belonging to different thread blocks but running on the same SM could also share data, typically only thread blocks that are neighbors in the thread block grid have overlap in their data footprints.
We consider the scheduling of neighboring thread blocks to the same SM out of about a hundred SMs as unlikely, because the assignment of thread blocks to SMs becomes increasingly incoherent and random as the kernel execution progresses.

The L1 cache allocates data at the granularity of 128B cache lines but transfers data at the granularity of 32B cache line sectors.
Therefore, a 128B granularity has to be used to compute the amount of allocated data in the L1 cache, but a transfer size of 32B for the computation of the transferred data volume.

The L1 cache uses a write-through policy, so that every store causes a transfer from the L1 to the L2 cache and has to be counted repeatedly.

The number of loads hitting on stored data in the L1 cache would depend on the order of the load and store operations, which is either difficult (done by the same thread) or impossible (load and store executed by different threads) to determine accurately.
Since numerical codes rarely have read-after-write sequences to the same memory location in the same kernel, we ignore this case in our estimator.
Such sequences would either be a missed optimization (happening inside the same thread), or a race condition (load and store from different threads).

The \pyth{CL32Visitor} function in listing \ref{lst:cl32visitor} can be used together with the generic grid iteration function in listing \ref{lst:griditeration} to count the number of unique 32B cache line sectors accessed by a representative group of threads.

We assume that fields do not alias each other, and compute a separate memory footprint for each field.
We replace the base pointer of the address expression, which is unknown at code generation time, by the alignment of the field.

\begin{figure}
\begin{python}
class CL32Visitor:
  CLCount = 0
  def count(addresses):
    # floor divide by 32 for 32B granularity
    # then count unique cache lines
    CLCount += len(unique(addresses // 32))
\end{python}
\caption {Simplified representation of a 32B cache line counting visitor}
\label{lst:cl32visitor}
\end{figure}

\subsection{DRAM - L2 Cache Data Transfer Volumes}

The L2 cache is a chip-wide shared resource, so that all threads share data in the L2 cache, even when they run on different SMs.
Because each SM can only run a limited amount of threads at the same time, depending on how many resources the threads require, only a comparatively small portion of the thread blocks in a kernel call grid are in execution at any given time.
New thread blocks are scheduled in $X-Y-Z$ order as older thread blocks complete, which results in a transient wave of running thread blocks with a sharp leading edge.
To simplify this, we subdivide the total thread block grid into discrete portions of thread blocks that can be resident on the SM simultaneously.
Such a discrete portion is commonly called a \emph{wave}, for example by NVIDIA's profiling tools.

Inside a wave, we consider all thread blocks to run simultaneously without any execution order.
Considering only sharing data inside a wave, the compulsory DRAM-L2 cache volume is the unique memory footprint of all threads of that wave.

Similar to the L1 cache, the L2 cache allocates 128B cache lines, but can transfer data at the granularity of 32B cache line sectors.

Different to the L1 cache, repeated stores are not directly written through to DRAM but get cached in the L2 cache, and are thus handled the same as load operations.

When partially written cache lines are evicted, the data of that cache line has to be read from the DRAM to complete the unwritten parts of the cache line.

\subsubsection{Implicit Thread and Address Sets}

Because the number of threads in a wave on modern GPUs is in the order $10^5$, we developed a method that does not require to explicitly enumerate every single thread coordinate to compute the unique memory footprint of a wave.
Instead, we use the Integer Set Library (ISL) \cite{ISL} to define and manipulate implicit sets of thread coordinates and memory addresses.
For example, the following notation defines a set of $[\mathsf{tidx}, \mathsf{tidy}]$ two dimensional thread coordinate tuples that contains all thread coordinates of the threads in a  $256\times 2$ thread block:

\begin{align*}
\mathsf{threads} := \{\:[\mathsf{tidx}, \mathsf{tidy}] :\; & 0 \leq \mathsf{tidx} < 256  \quad \land & \\ \nonumber
                                                                        & 0 \leq \mathsf{tidy} < 2 \qquad \}
\end{align*}

For the DRAM volume estimation we use an implicit description of a set of all thread coordinates of a representative wave in the middle of the call grid.
The ISL has functions to manipulate these sets, like intersecting this wave thread set with the set of valid domain coordinates, which accounts for the very common \lstinline[style=C++]{if(threadIdx.x > width-1)  return;} pattern necessary for thread block sizes that are not clean dividers of the valid iteration domain.

Each memory access is described by a mapping from a thread coordinate to the memory address it references.
A $\operatorname{floor}$ divide by the cache line size, in this case 32B, makes the mapping map to cache lines instead of addresses.
For example, the memory reference \lstinline[style=C++] {A[tidx][tidy+1]} to a two-dimensional, $100$-wide array where the base pointer has an alignment shifted by one element (-1) would be represented as:

\begin{align*}
\{[\mathsf{tidx, tidy}] \rightarrow [\mathsf{idx}] : & \\ \nonumber
 \qquad \mathsf{idx} = \operatorname{floor}  ( (-1 + & \mathsf{tidx} + (\mathsf{tidy+1}) * 100  ) * 8 / 32 )\}
\end{align*}

Applying that mapping to the set of thread coordinates results in a set of accessed cache lines.
The union of the address sets of all memory accesses results in the unique memory footprint of the representative wave.
Using the function \pyth{count_val} to compute the cardinality of this set gives the information about the number of cache lines in the unique memory footprint.

For addressing into multidimensional arrays, the computation of a linear, one dimensional address from the multidimensional cell coordinates can result in complex set descriptions, because the referenced memory addresses form non-contiguous, scattered ranges.
To keep the set descriptions simple, which ensures that the function computing the number of elements in a set evaluates quickly, we optionally skip the necessity to convert multi dimensional array references to a one dimensional, linear addresses, but instead support a multi dimensional address space.
Two multi dimensional addresses are considered distinct, when their address tuples differ in at least one element.
This would not be exactly true for two array coordinates, if a coordinate wraps around in one dimension. For example, the two coordinate tuples $(150, 1)$ and $(50, 2)$ in a $100\times100$ 2D array would be considered distinct in the 2D address space, but would actually be the same in the linear array address space, because the X coordinate has wrapped around.
We assume that codes usually do not do that on purpose, though.

The memory reference \lstinline[style=C++] {A[tidx][tidy+1]} from before would instead be represented by this mapping from a 2D thread coordinate to a 2D address:

\begin{align*}
\{[\mathsf{tidx, tidy}] \rightarrow [\mathsf{ax}, \mathsf{ay}] : \: & \mathsf{ax} = \operatorname{floor}  ( \mathsf{tidx} * 8 / 32 ) \quad \land \\
  & \mathsf{ay} = \mathsf{tidy+1}) \}
\end{align*}

The innermost, X dimension still contains the floor division by the cache line size to account for the fetch granularity.

The counting of cache lines in a multidimensional address space differs in two ways from counting in a linear address space.
In the case where an element at the boundary of an array is in a cache line that wraps around to the next row, a separate cache line would be counted at the beginning and the end of a row.
Furthermore, the alignment of arrays cannot be considered in multidimensional addressing any more, because every row could have a different alignment.
However, we found that these inaccuracies amount to minuscule differences for larger grids.

Overall, the big advantage of implicit set descriptions is the decoupling of the evaluation runtime from the number of threads in the grid, which is good for a device wide metric like the DRAM data volume where large thread grids have to be considered. The ISL also enables the computation of advanced set relationships, like the intersection, subtraction and union of address sets.

\subsubsection{Warm Cache Reuse}

\begin{figure}
  \centering
  \includegraphics[width=\columnwidth]{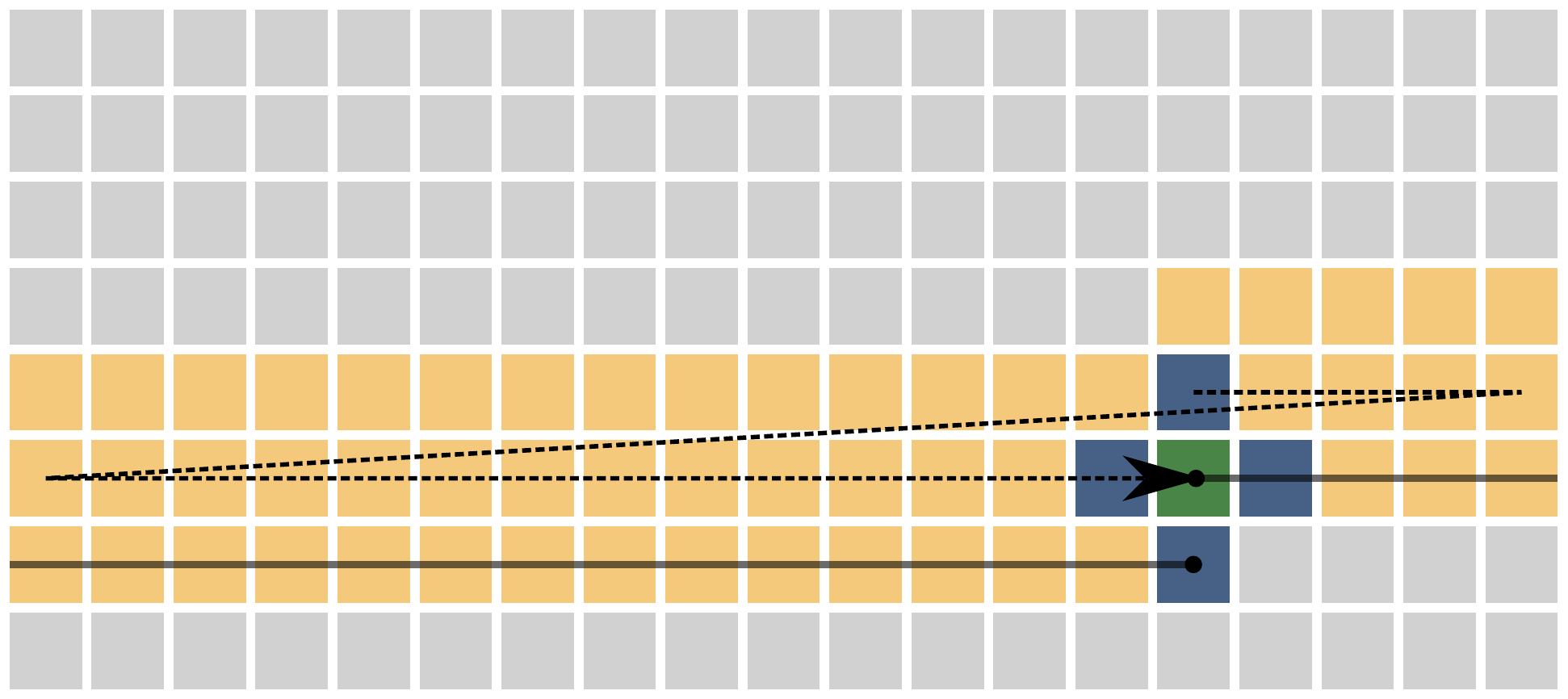}
  \caption{Illustration of the layer condition for a 2D 5-point stencil. The dashed line shows the iteration order.}
  \label{fig:layer_condition}
\end{figure}

A wave's data is not gone after it finishes executing, and so threads can hit on data that previous waves have used, if the data has not been evicted due to capacity yet.

In \cite{SBAC}, we have shown how intersecting the current wave's L2 footprint with the directly preceding wave's footprint allows to determine the data volume that can be saved due the cache being warmed up by previously running threads.
On the V100, the relatively small amount of L2 cache capacity per-thread limits of how long data stays available for reuse after a thread has finished, so that there was no need to look further into the past than two waves.
The A100 has about $2.5\times$ more per-thread L2 cache capacity than the V100, which also increases the chance for reuse between threads that are farther apart.

Approaches like \cite{reuse} use the concept of \emph{reuse distance} to determine whether a previously loaded datum is still available for reuse.
However, there are two reasons why this approach is not suitable for our purposes.

For one, reuse distance relies on a sequential order of memory operations, which does not exist on GPUs due to their parallel nature.
Even though thread blocks are scheduled in a clear sequential order, there is no deterministic order in which operations are executed.
We simplify this blurred sequentiality by only differentiating whether a operation happens before or after a point in time.
For example, all operations of the current wave happen simultaneously, in no assumed order, but only after all operations of previous waves are completed.

Secondly, a precise estimate using reuse distances would require iterating through all memory operations until the reuse distance becomes too large, which conflicts with our goal of a quickly evaluable estimate.

\begin{figure}
  \centering
  \includegraphics[width=0.9\columnwidth]{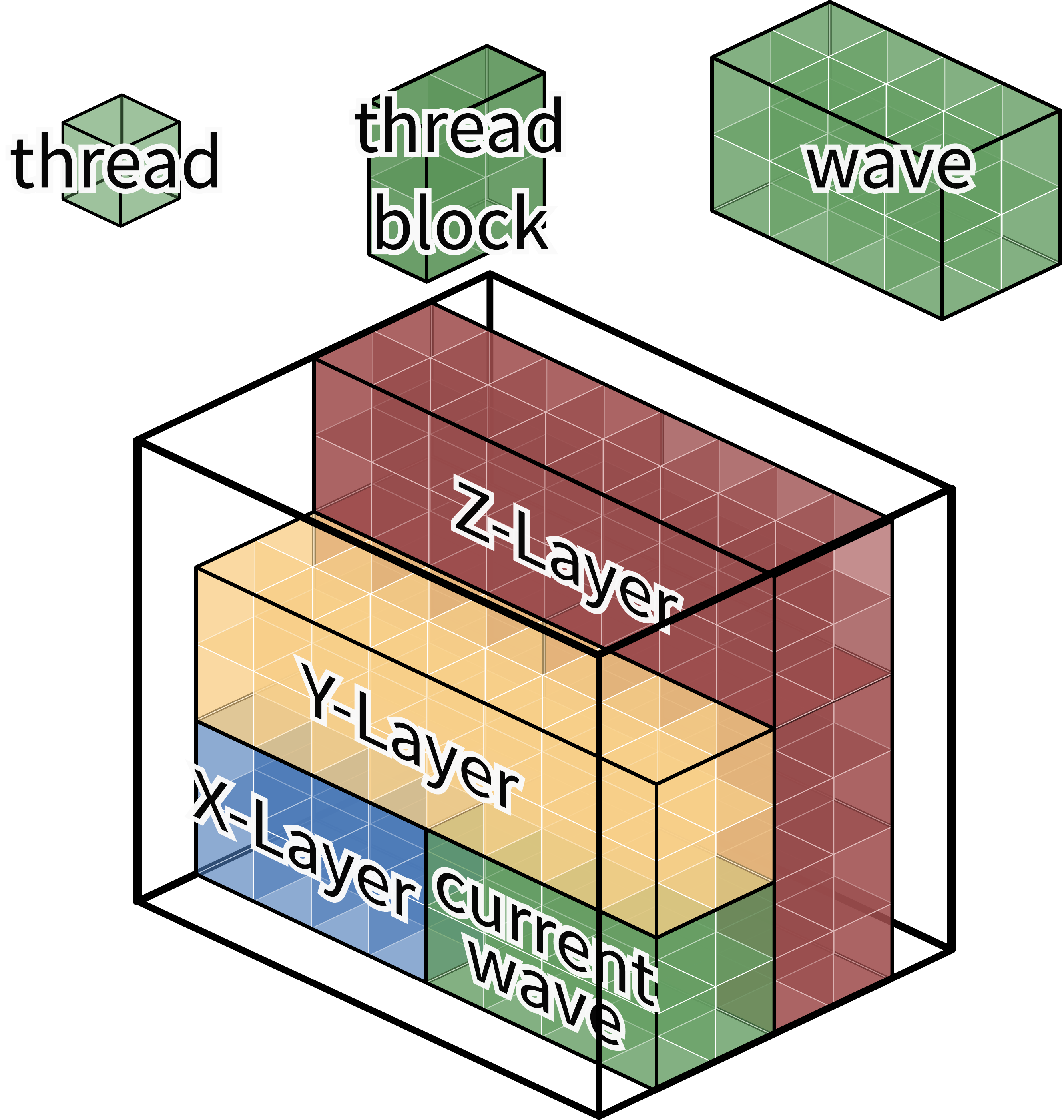}
  \caption{Illustration of how the layer condition thread sets are constructed. Each thread block has $1\times2\times2$ threads, and a wave comprises 4 thread blocks. For better visibility, thread blocks back to front.}
  \label{fig:thread_sets}
\end{figure}

Instead, we exploit that for the sequential traversal of structured grids, there are typical reuse distances for spatial reuse.
For example, figure \ref{fig:layer_condition} illustrates that the reuse distance from the bottom to the central value of a 2D 5-point stencil is one row.
For reuse in the z dimension, the distance would be one xy layer.
A cache level has to be large enough to keep all referenced data between the two uses, which is what the so called \emph{layer condition}, described in \cite{layercondition1} and \cite{layercondition2}, is based on.
For sequential grid iteration, the 3D layer condition $3 \cdot x \cdot y \cdot 8B < V_{L2} / 2$ would require three layers of the grid to fit into the cache capacity, half of which is available for the source grid.

We transfer the concept of layer conditions, which is valid for sequentially iterated grids, to our case of parallel grid iteration by looking at sets of concurrently executing threads.
We define a set of threads for each dimension, illustrated in figure \ref{fig:thread_sets}, that is adjacent to the current wave in that dimension and contains the threads with potential spatial reuse in that dimension.
The threads of that set are one reuse distance in that dimension away from the threads of the current wave.
The intersection of the unique memory footprint of each set and the one of the current wave is the amount of potential reuse along this dimension, and the data volume of each set's unique memory footprint decides whether the layer condition is fulfilled.

For many thread block and grid sizes, the X and sometimes the Y dimension is already filled by the current wave, in which case these sets are empty.

\subsection{Capacity Misses}

\begin{figure}
  \centering
  \includegraphics[width=\columnwidth]{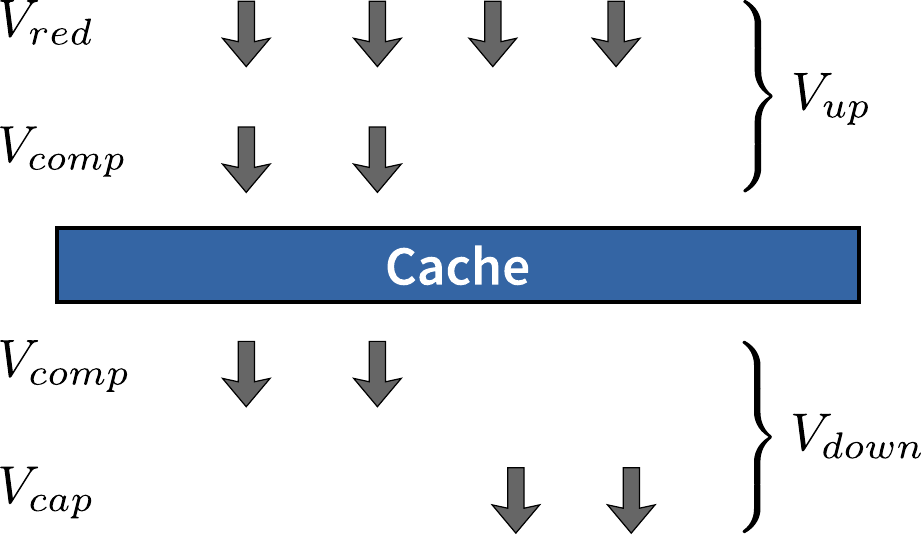}
  \caption{Illustration of in- and outgoing data volumes on a cache hierarchy level}
  \label{fig:capacity}
\end{figure}

Besides the \emph{compulsory} cache misses, the classification of cache misses by \cite{hennessy-patterson} also recognizes  \emph{capacity} and \emph{conflict} cache misses.
Capacity misses are cache misses that miss because the capacity of a cache level is not sufficient to keep the required data volume.
Conflict misses happen when data is being evicted earlier than necessary because cache aliasing reduces the effective cache capacity.
Conflict misses are not considered in this paper as a phenomenon separate from the capacity misses, as the occurrence of aliasing is too dependent on the details of an architecture's cache organization and also highly volatile.

The data volume transferred due to capacity cache misses is a percentage of the \emph{redundant} data volume, which is the volume that has the potential to hit in a cache level.
We denote the total volume of data transferred by a cache to the memory hierarchy level above it as $V_{up}$, and the volume of data caused by requests of that cache level to the level below it as $V_{down}$ (see figure \ref{fig:capacity}).

For memory hierarchy levels $N$ and $N-1$ (the latter being closer to the registers), we have $V_{up}^{N} = V_{down}^{N-1}$.
$V_{down}$ can be split into three components:

\begin{equation}
  V_{down} = V_{comp} + V^{*}_{cap} + V^{*}_{conflict}
\end{equation}

$V^{*}_{cap}$ and $V^{*}_{conflict}$ would be the pure capacity and conflict misses, but are difficult to observe separately, and are replaced by an observed capacity volume $V_{cap} = V^{*}_{cap} + V^{*}_{conflict}$.

 The total data volume $V_{up}$ between a cache level and the memory hierarchy level above can be split up into the compulsory volume $V_{comp}$, which consists of the data accessed for the first time by that collaborative group, and the redundant volume $V_{red}$ comprising repeated requests for the same data:

\begin{equation}\label{eq:vup}
  V_{up} = V_{comp} + V_{red}
\end{equation}

While the order of memory accesses inside a thread is fixed, the memory access order among different warps is not deterministic.
It is therefore impossible to analytically compute the exact amount of capacity misses, but it is possible to identify scenarios where capacity misses happen at all and to estimate the extent.

We define the capacity hit and miss rates $R_{hit}$  and $R_{miss}$ as the portion of redundant accesses that will hit or miss in the cache:

\begin{equation}\label{eq:vred}
  R_{hit} = 1 - R_{miss} = 1 - \frac{V_{cap}}{V_{red}}
\end{equation}

Note that $R_{hit}$ is not the hit rate of the entire cache, but only the hit rate in the redundant data volume.
The complete cache hit rate would be lower, because it additionally has the compulsory volume in the denominator.

We assume that the major factor that influences this portion is the data volume $V_{alloc}$ allocated by the threads sharing that cache level.
The allocation volume can be different from $V_{cap}$, for example in the case of Volta's L1 cache that uses a 128B for cache line allocation, but a 32B granularity for data transfers.
We define the oversubscription factor $O$ of that cache level as the ratio of the available cache capacity $V_{cache}$ and the allocation volume:

\begin{equation}
O = \frac{V_{alloc}}{V_{cache}}
\end{equation}

It is impossible to give an analytic formula $R_{hit}(O)$ for even a single application, but the behavior of that relationship in the limit can be characterized.
For an oversubscription factor smaller than one, there is enough cache capacity for the complete footprint and $R_{hit}$ should be one.
With increasing oversubscription, $R_{cap}$ should go towards zero.

As a stand-in for a smooth transition between the two states we fit a sigmoid function of the form $\hat R_{hit}(O) = a e^{-b e^{-c\,O}}$.
This function is not grounded in any real world mechanism, and other functions that smoothly interpolate between the two states could be used.

By rearranging equation \ref{eq:vred} and using $\hat R_{hit}(O)$ as a hit rate esimate, the capacity miss volume is then the product of the capacity miss rate and the redundant volume, which can be computed from equation \ref{eq:vup} as the difference between $V_{up}$ and $V_{comp}$:

\begin{align}
  V_{cap} = & (1 - \hat R_{hit}(O))\: V_{red}  \nonumber \\
         = & (1-\hat R_{hit}(O)) (V_{up} - V_{comp})
\end{align}

$V_{up}$ and $V_{comp}$ are both volumes for which estimates already exist.

We use separate hit rate fit functions for the redundant loads in the L1 cache ($\hat R_{hit}^{L1}$), for the  overlapping volumes in the y and z layers ($\hat R_{hit}^{L2,over,y}$ and $\hat R_{hit}^{L2,over,z}$), and for redundant stores in the L2 cache ($ \hat R_{hit}^{L2,store}$).

\section{Evaluation}

\begin{figure}
  \centering
  \includegraphics[width=\columnwidth]{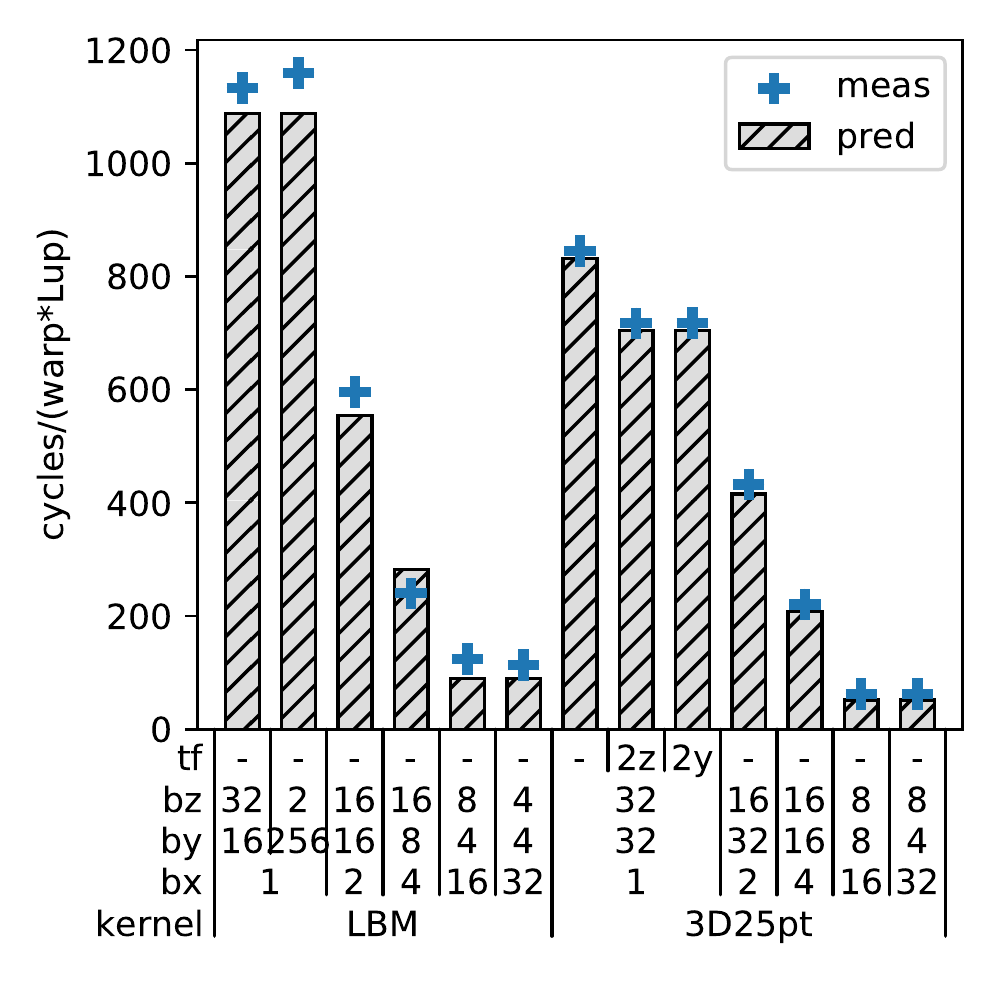}
  \caption{Cycles to complete all L1 cache transfers of one warp wide lattice update. Estimated cycle count vs hardware performance counter \lstinline[style=C++]{l1tex__data_pipe_lsu_wavefronts}.}
  \label{fig:l1thru}
\end{figure}

\subsection{Data Points}

The accuracy of the metric estimator and the performance model is evaluated by comparing
predictions and estimates for a range of different thread block sizes  $(X,Y,Z)$ that fulfill:

\begin{align}
  & X,Y \in \{1,2,4,8,16,32,64,128,256,512,1024\}  &\land \\ \nonumber
                 & Z \in \{1,2,4,8,16,32,64\} &\land \\ \nonumber
                 & X\times Y \times Z = \begin{cases} 1024 : stencil \\ 512 : LBM \end{cases} &
\end{align}

The selection is limited to thread block sizes with the same total number of threads to avoid overcrowding of the graphs. The register requirements for the LBM kernel necessitates a thread block of at most 512 threads to avoid spilling.

\definecolor{myred}{rgb}{1,0.1,0.0}
\definecolor{mygreen}{rgb}{0.25,1.0,0}
\definecolor{myblue}{rgb}{0.25,0.25,1.0}
\definecolor{myyellow}{rgb}{1.0,0.9,0.0}
\definecolor{mypurple}{rgb}{0.625,0.0,1.0}
\definecolor{myturq}{rgb}{0.0,0.625,1.0}
\definecolor{mygray}{rgb}{0.55,0.5,0.5}

The plots use color coded markers to differentiate different thread block sizes by mapping the three thread block coordinates to RGB colors: $RGB(\frac{log \: x}{log \:256}, \frac{log \:y }{log \:256}, \frac{log \:z }{ log \:64})$. This means that the aspect ratios of thread block sizes have typical colors: Shapes with only one long dimension are colored in the primary colors \fcolorbox{white}{myred}{red} (wide), \fcolorbox{white}{mygreen}{green} (tall) and \fcolorbox{white}{myblue}{blue} (deep). Shapes that are short in one dimension feature mixed colors like \fcolorbox{white}{myyellow}{yellow/orange} (wide \& tall, but shallow), \fcolorbox{white}{mypurple}{purple} (deep \& wide, but squat) or \fcolorbox{white}{myturq}{turquoise} (tall \& deep, but narrow). Cubish shapes will have desaturated \fcolorbox{white}{mygray}{gray/beige} colors.

\subsection{Long Range 3D Stencil}

The first application is a range four 3D 25-point star stencil.
At 25 floating point operations and a minimum of one double precision grid point or 8B loaded and stored per lattice update (LUP), the arithmetic intensity of $~1.5 \frac{\mathsf{Flop}}{\mathsf{B}}$, is below the machine balance of $3.5 \frac{\mathsf{Flop}}{\mathsf{B}}$ of a A100 for that instruction mix, which still makes this stencil memory bound.

We use a straight forward 3D iteration scheme for the stencil, combined with \textit{thread folding}, a technique where each thread computes multiple consecutive grid coordinates. We test each block size with $2 \times$ thread folding in $y$ or $z$ direction as well as no thread folding.

We find that on modern GPU architectures like Volta or Ampere, other, more complex iteration schemes and optimizations for stencils, which try to deal with limitations like the comparatively small cache capacity, are not as relevant any more, as these architectures feature much larger caches than previous GPUs.

For example, \cite{higherorderstencils} finds the straight forward 3D mapping to be the fastest on a Volta GPU, if the thread block sizes are chosen well.

The grid size for the stencil is $640\times512\times512$.
This grid size has purposefully been chosen at a transition point for the 3D layer condition.
For sequential traversal, the 3D layer condition $640 \cdot 512 \cdot 8B \cdot 3 = 7.5MB < V_{L2} = 20 MB$ would be fulfilled.
This rule is also indicative for waves that consist only of one z layer of threads.
For thread block sizes with a z extent larger than one, the wave shape is correspondingly deeper, and more layers need to be fitted into the cache.

For the fitting of the hit rate functions $\hat R_{hit}(O)$, a wide range of domain sizes has been used to make sure that all scenarios are covered.

\begin{figure}
  \centering
  \includegraphics[width=\columnwidth]{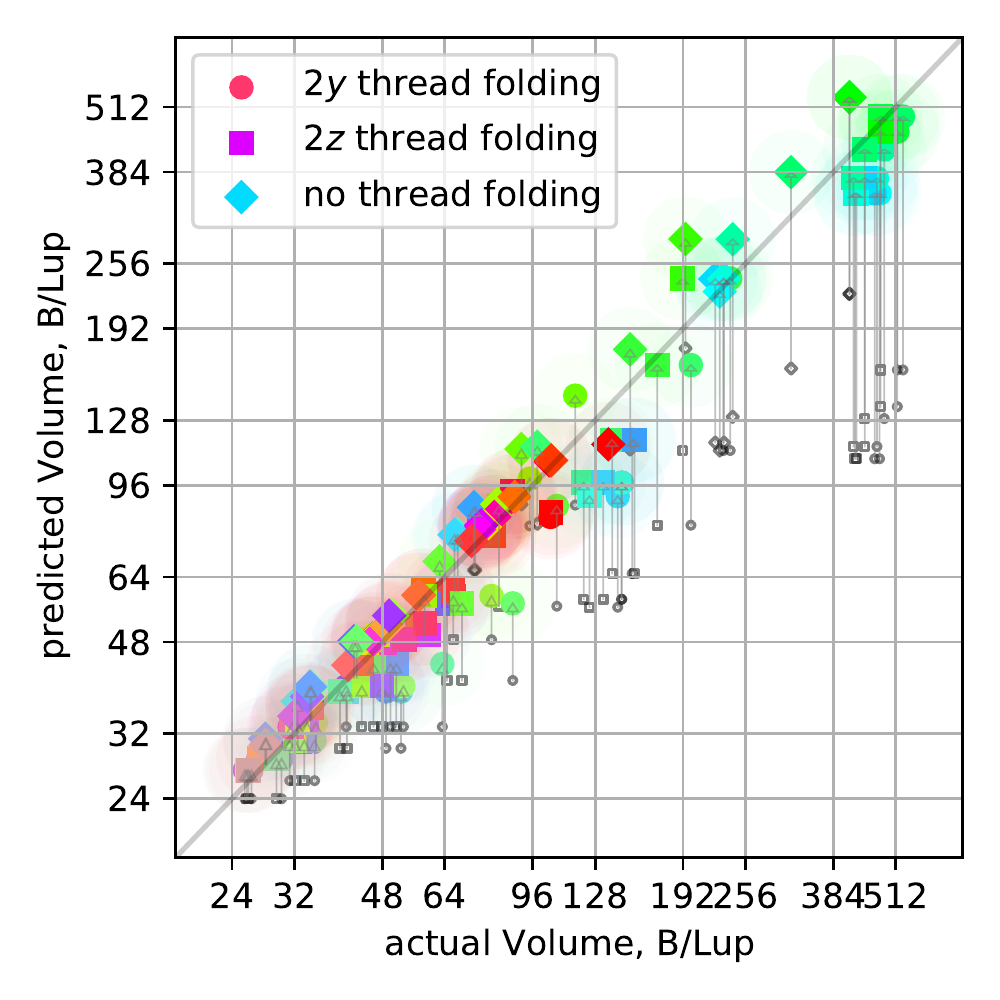}
  \caption{3D25pt/range 4 star stencil: predicted vs measured L2-L1 loaded data volume. Data points colored by block size. Gray markers and arrows show comparison without modeling capacity misses.}
  \label{fig:l2loadstencil}
\end{figure}

\subsection{Multi Phase LBM}

The second application is a LBM kernel using the conservative Allen-Cahn model, which can solve highly complex two-phase-flow phenomena.

We will focus on the kernel doing the interface tracking, which is one of two LB schemes that this method consists of. The curvature of the phase field is computed with a finite difference discretization, which adds a 3D7pt stencil to the conventional, very memory intensive D3Q15 LBM stencil.
This makes it particularly complicated to achieve good performance results \cite{lbmpy2}, but at the same time interesting for automatic performance estimations, which can give significant insight into the complex structure of the compute kernels.

Each lattice cell is resolved with 15 probability distribution function (PDF) values for the LB calculation, and additionally, the information of the phase-field of 6 neighboring lattice cells is needed. For propagating the PDF values to the neighboring lattice cells, a pull scheme is used. Thus the stores are aligned, and the loads are not. The D3Q15 stencil LBM part of the kernel transfers a data volume of $2*15*8B = 240B/Lup$ read and write without any reuse, compared to the just $16 - 64B/Lup$, depending on the cache effectiveness, for the finite-difference stencil.

\subsection{L1 Load Volume Prediction}

\begin{figure}
  \includegraphics[width=\columnwidth]{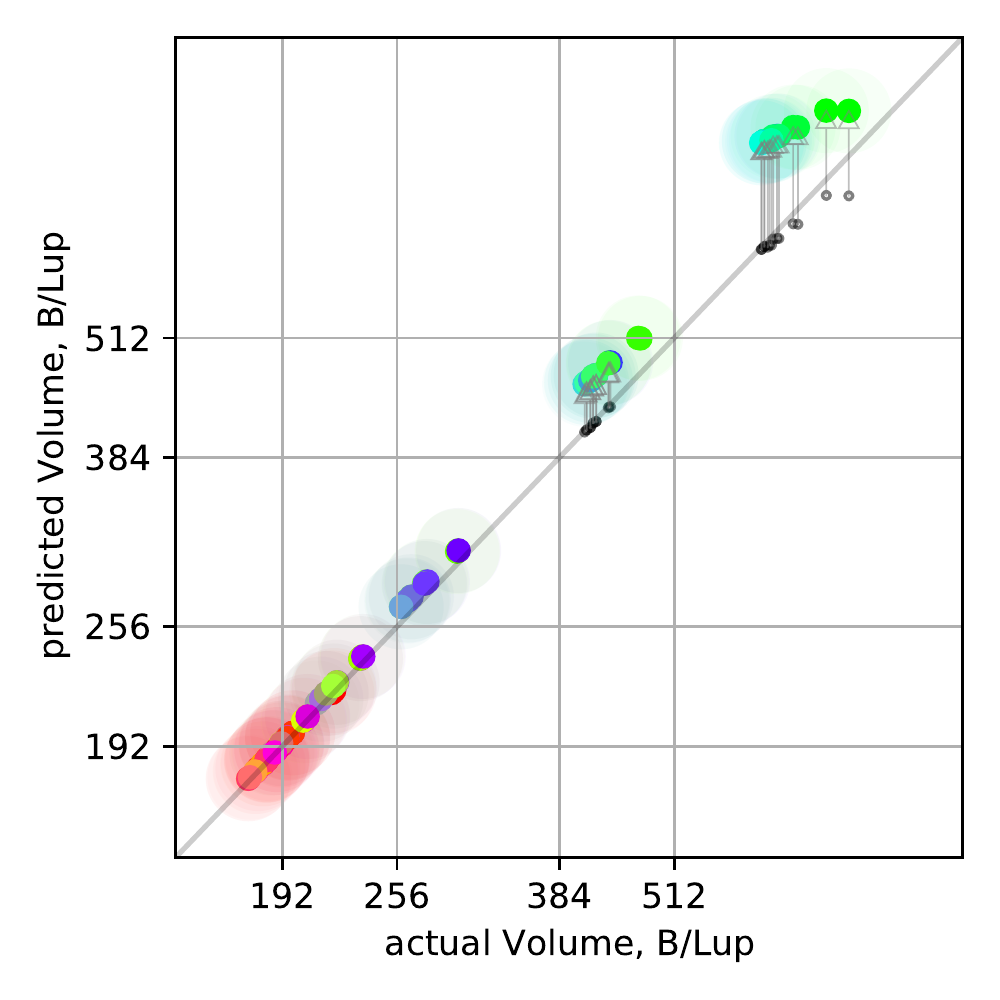}
  \caption{LBM Kernel: predicted vs measured L2-L1 loaded data volume. Data points colored by block size. Gray markers and arrows show comparison to estimate without modeling capacity misses.}
  \label{fig:l2loadlbm}
\end{figure}

\begin{figure*}
  \centering
  \includegraphics[width=\textwidth]{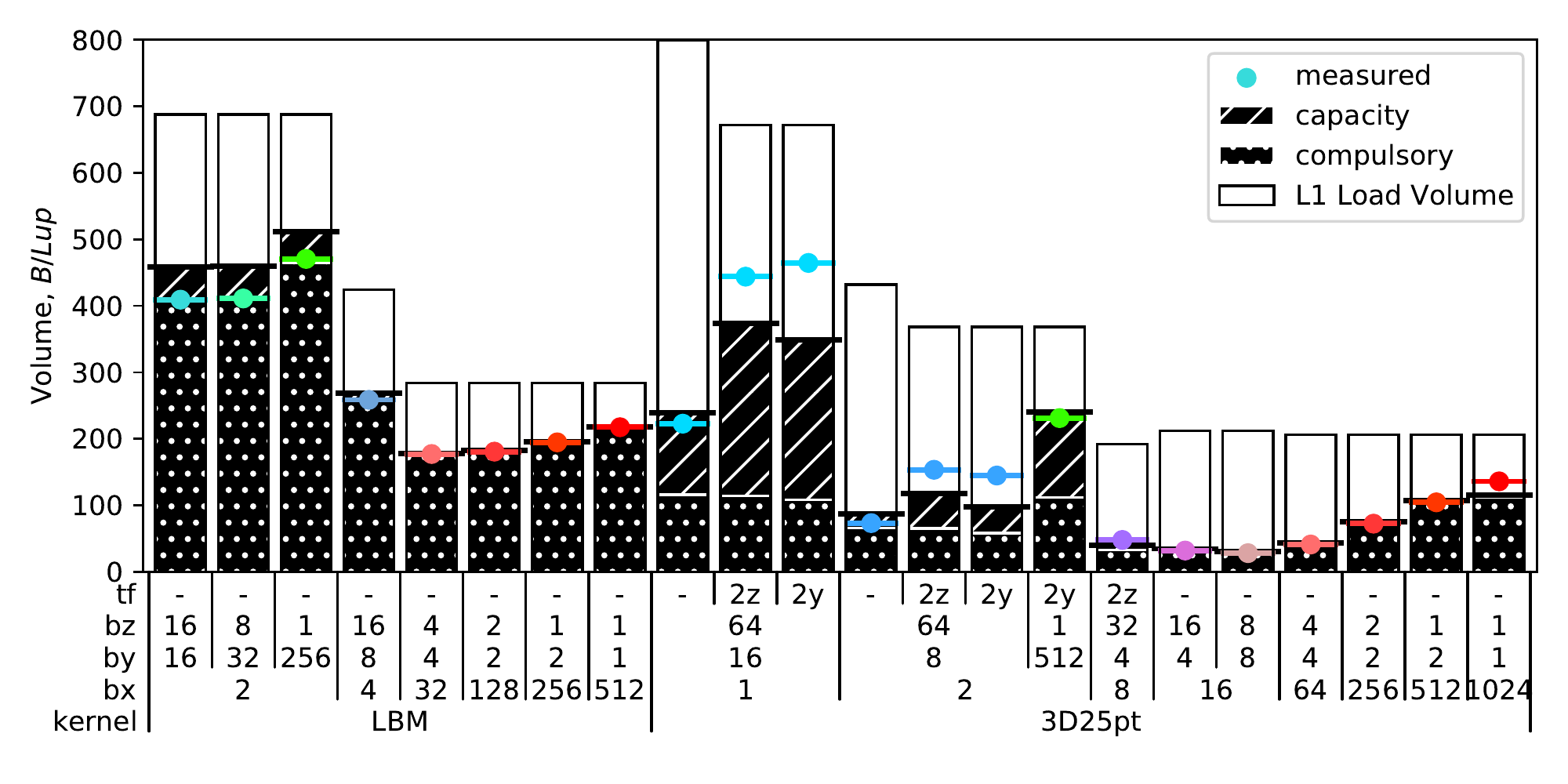}
  \caption{Composition of the L2-L1 load data volume for selected block sizes}
  \label{fig:l2comp}
\end{figure*}


Figure \ref{fig:l1thru} shows the prediction of how many cycles each warp occupies an SM's L1 cache for a single lattice update.
For the access patterns of the two kernels, the thread block width is also the width of the contiguous blocks of accessed memory.
For 16 threads or higher, each half-warp of 16 threads loads a single contiguous block without any bank conflicts, that can be processed in a single wavefront.
With each decrease in the thread block width, the number of bank conflicts increases.
The L1 time per lattice update can be decreased through thread folding, because values can be reused from registers instead of reloading them.
This can be seen for the data points with the \emph{2y} or \emph{2z} label, which have a lower L1 cache utilization than the data point with the same thread block width but without thread folding.

The comparison with the measured number of L1 cache wavefronts, as recorded by the hardware performance counter \lstinline[breaklines=true,style=C++]{l1tex__data_pipe_lsu_wavefronts}, shows that the estimation of the L1 cache throughput is highly accurate.

\subsection{L2-L1 Data Volume Prediction}

Figures \ref{fig:l2loadstencil} and \ref{fig:l2loadlbm} plot the estimate of the L2-L1 cache load data volume in comparison to the actual data volume measured using the hardware performance counter \lstinline[style=C++]{lts__t_sectors_srcunit_tex_op_read}.
Plot points on the diagonal would be block sizes where the estimate is identical to the actual value, whereas points in the upper-left / lower-right triangular area of the plot are over / under estimated.

In figure \ref{fig:l2comp}, these predictions are broken down into compulsory and capacity reason for a selected number of thread block sizes.
The shown L1 cache load volume (black outlined bar) is an upper limit on the L2-L1 volume, and would be completely filled by the capacity miss bar if $R_{hit} = 0$, and empty if $R_{hit} = 1$.

For the stencil, the different thread block sizes cause a wide range of L2-L1 cache data volumes.
Pale, low saturation colors, representing thread block sizes with balanced, cube shaped dimensions, e.g. $(16,8,8)$, occupy the low volume  lower left corner of figure \ref{fig:l2loadstencil} due to their low surface volume ratio.
For the same reason, thread block sizes with one short dimension, e.g. $(512,2,1)$ or $(2, 512, 1)$ in red and green, have higher data volumes.
A very short x dimension e.g. $(1, 16, 64)$, represented with colors without red, i.e. green and blue, lead to both inefficient use of the transferred 32B cache line sectors, and also inefficient allocation of 128B cache lines and consequently, large amount of capacity misses.

Thread folding in the right dimension leads to lowered compulsory volumes, but also to increased capacity misses.
The majority of the data points that have capacity misses are using thread folding.

The difference in the compulsory volumes is less pronounced for the LBM kernel, due to the high amount of streaming data volume.
The streaming component favors the $x$ dimension to be as large as possible (which corresponds to red/orange/purple colors), as this minimizes the percentage of partially used cache lines caused by unaligned loads.
As a compromise with the preference of the stencil part for cubed shapes, thread block sizes like $(32,4,4)$ show the lowest volumes here.
The amount of capacity misses is generally very low.
We assume that the non-LRU replacement policy of Volta's/Ampere's L1 cache replaces data with streaming access patterns preferentially, so that the data of the stencil component is not displaced by the large data volume loaded by the LBM part.

As a future extension, it would be possible to automatically classify allocated data volume into streaming, i.e. a single reference, and reuse volume to differentiate the two scenarios.


\subsection{DRAM-L2 Data Volume Prediction}

\begin{figure*}[t]
  \centering
  \begin{minipage}[t]{.32\textwidth}
  \includegraphics[width=\textwidth]{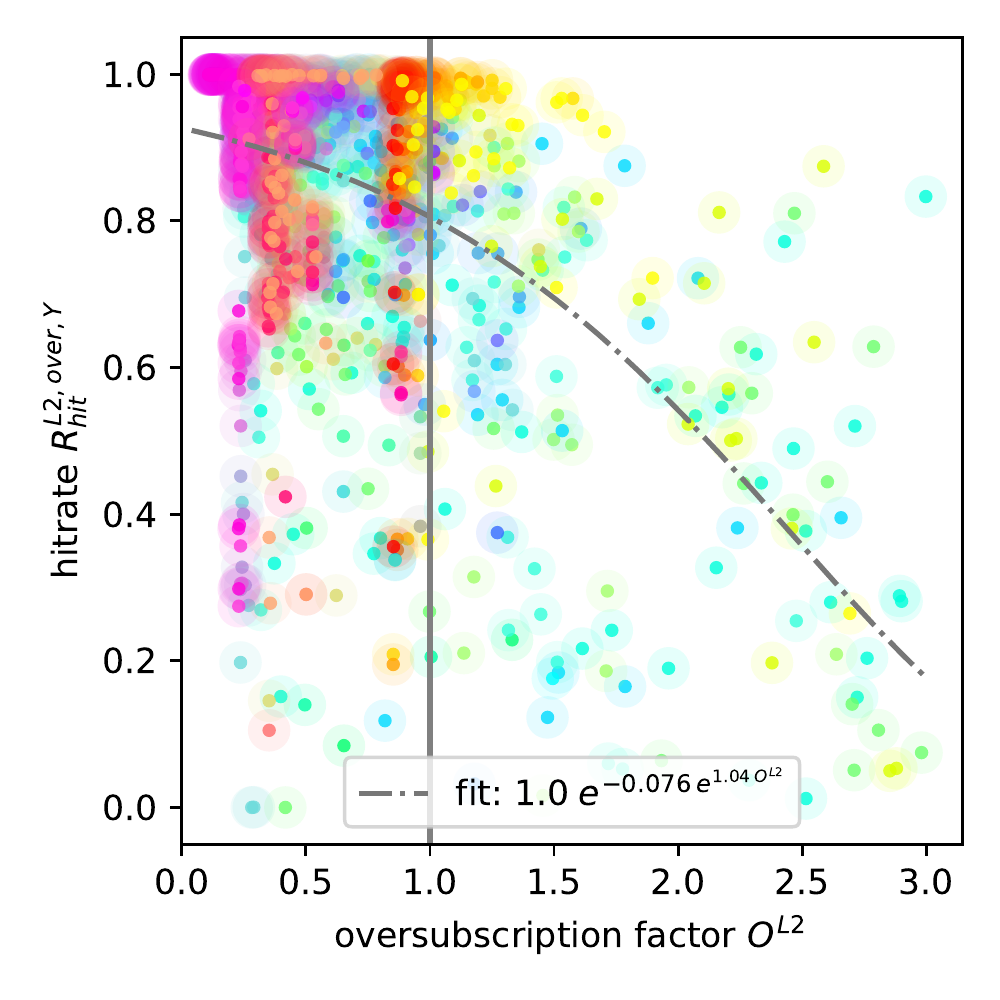}
  \caption{L2 cache oversubscription vs hit rate in the y-layer overlapping volume and the hit rate fit function }
  \label{fig:rovery}
\end{minipage}
  \hfill
\begin{minipage}[t]{.32\textwidth}
  \centering
  \includegraphics[width=\textwidth]{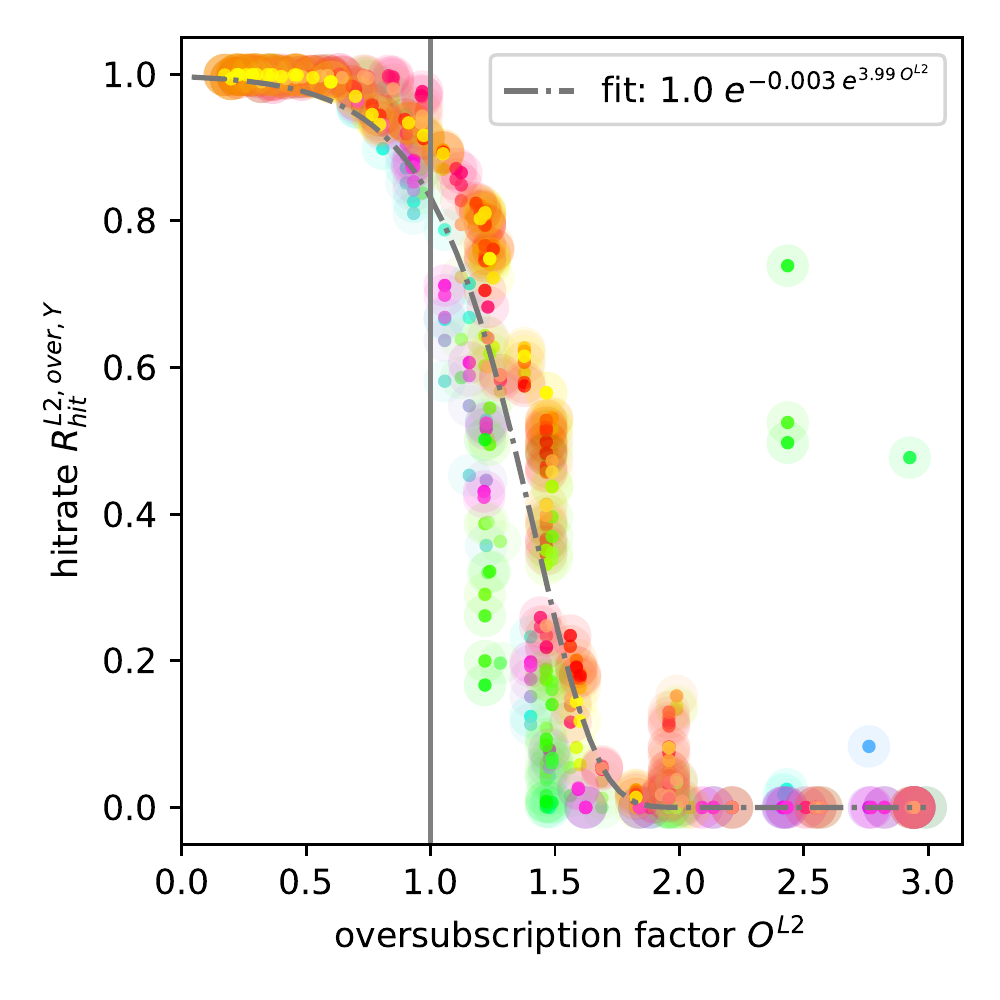}
\caption{L2 cache oversubscription vs hit rate in the z-layer overlapping volume and the hit rate fit function}
  \label{fig:roverz}
\end{minipage}
  \hfill
\begin{minipage}[t]{0.32\textwidth}
  \includegraphics[width=\textwidth]{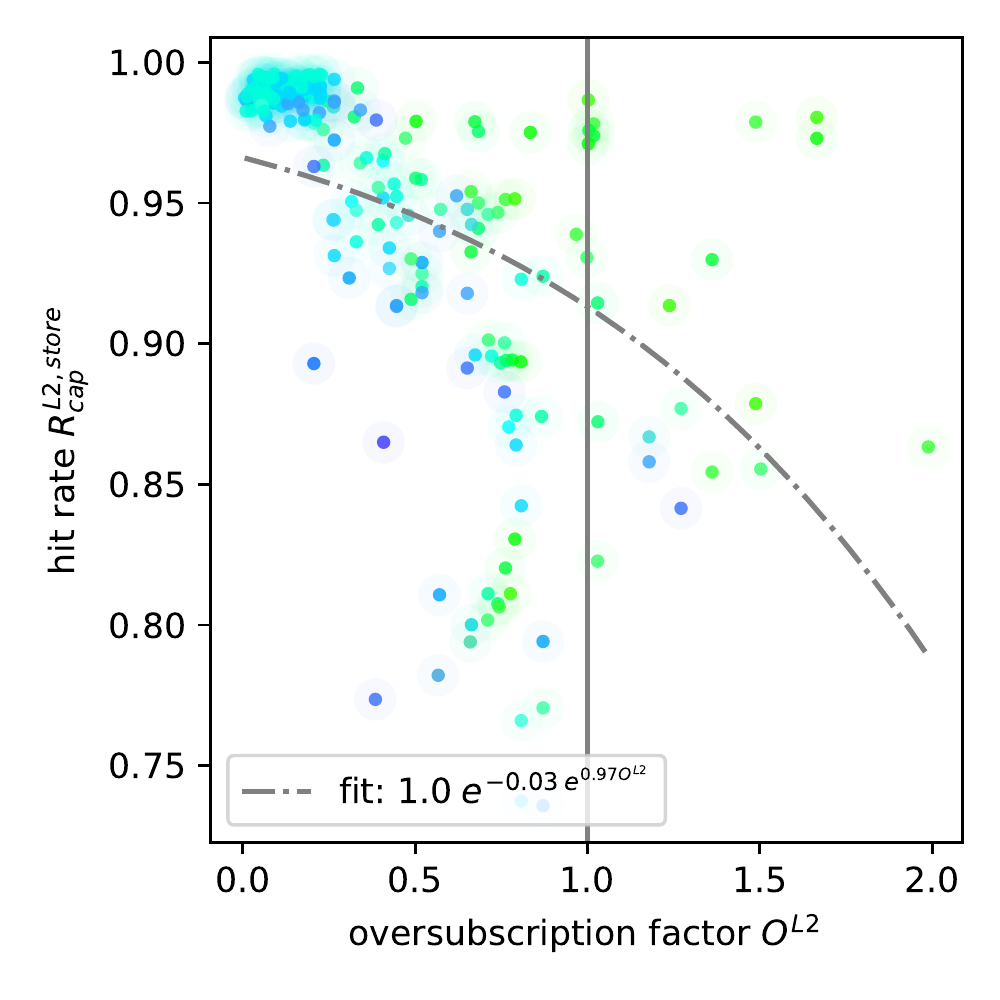}
  \caption{L2 cache oversubscription vs hit rate for redundant stores and the hit rate fit function}
  \label{fig:rstoremiss}
\end{minipage}
\end{figure*}

The hit rate ratio fit functions $\hat R_{hit}^{L2,over,y}$ and $\hat R_{hit}^{L2,over,z}$ for the hit rates in the overlapping layer volumes are plotted in figures \ref{fig:rovery} and \ref{fig:roverz}.
The hit rates to be fitted can be determined by deducting the minimal compulsive data volume excluding the overlapping data volume from the measured data volume and dividing the difference by the overlapping volume.
For both curves, a wide range of domain sizes and thread block sizes has been used.

For most data points, the y layer volume in figure \ref{fig:rovery} stays well within the capacity of the L2 cache, and the average hit rates stay high.
Only for domains with large xy planes larger than $10000\times1000$ does this become a factor.
The y layer overlap hit rates depend only loosely on the oversubscription factor, with wide variations in the actual overlap hit rates.
One reason is that in reality, the execution order of thread blocks is less strict than our model, and the thread blocks belonging to the current wave and the ones that belong to the y layer overlapping thread do intermingle, because of the frayed tail end of the wave.

Another reason is the different thread scheduling order for different thread block sizes.
The thread block sizes $(512,1,1)$ and $(1,512,1)$ result in the same wave and overlapping y layer thread sets, but the order in which threads are spawned inside the wave is different.
In the first case of the wide thread block shape, the wave is filled top to bottom, with all the threads that have potential y layer reuse spawning first.
For the tall shape, the wave fills left to right, which has more potential of evicting other data in the meantime.
This is a case where the model assumption of simultaneous execution of thread blocks without ordering inside a wave misses some effects.

The curve is much tighter and steeper for the z layer hit rate in figure \ref{fig:roverz}.
Inside the L2 cache capacity, for oversubscription factors smaller than one, the z layer overlap hit rate is close to one and then quickly drops 0 for oversubscription factors more than two.
The deviations are much smaller, because the threads in the z layer overlap thread set are much further in the past and more clearly separated from the set of threads currently executing.
The difference in intra wave thread scheduling due to different thread block sizes is again visible, as wide thread block shapes in red are generally above the fit curve, and the tall shapes in green below the curve.

The miss rate for redundant stores is plotted in figure \ref{fig:rstoremiss}.
In the examined scenarios, redundant stores only occur for the case of stores that write partial cache lines.
Because all stores are aligned, this in turn only happens for thread block sizes with width one or two, which only write one or two values of a four DP elements wide cache line sector.
The allocated volume fits into the L2 cache in most cases, which is why the hit rate is over $90\%$ for most data points.
Applications with larger allocated L2 volumes would be required to get data points for higher L2 cache oversubscription rates.

\begin{figure*}
  \centering
  \includegraphics[width=\textwidth]{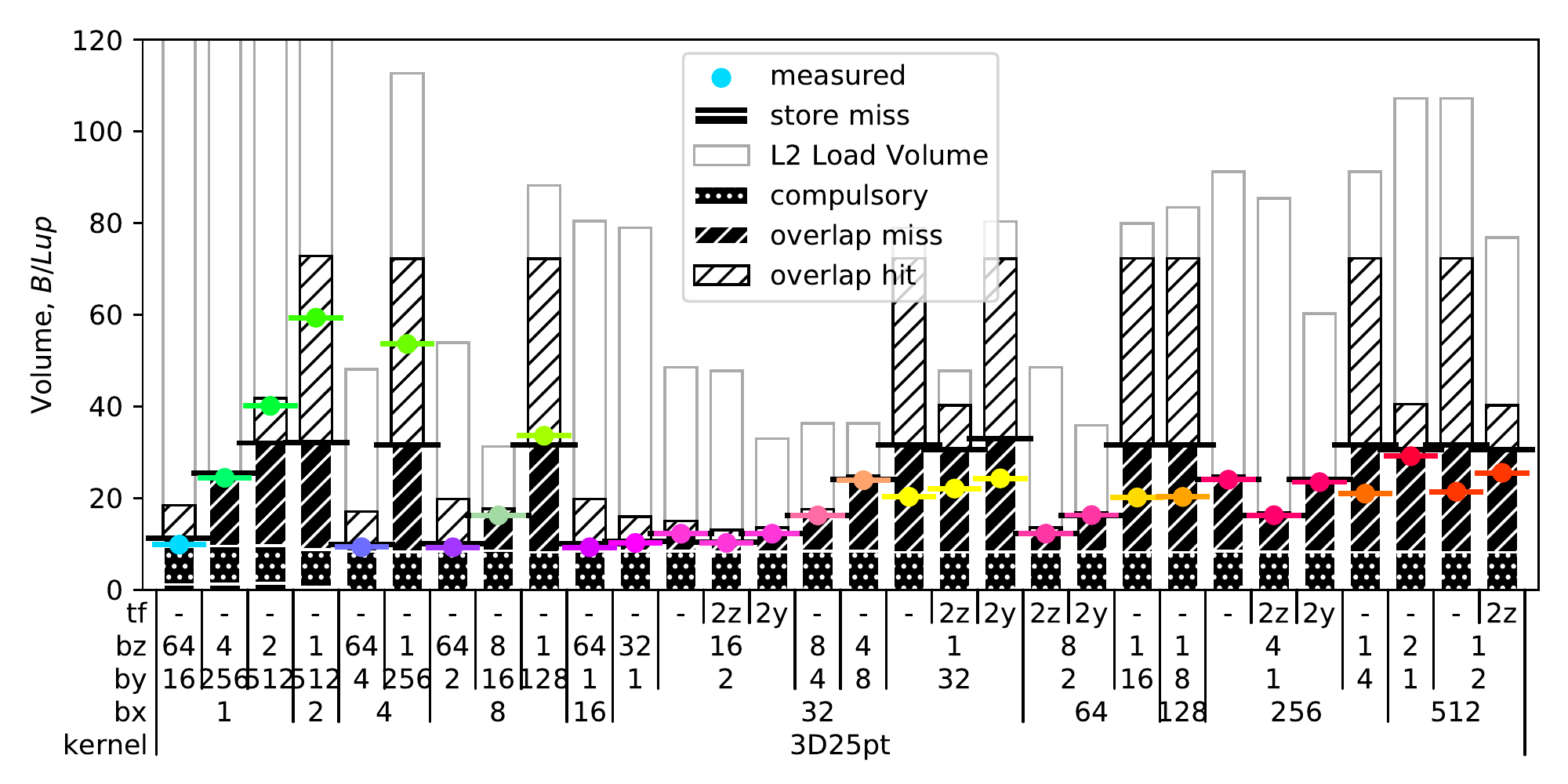}
\caption{Composition of stencil DRAM load data volumes for selected block sizes $(bx, by, bz)$ and thread folding factors ($2y$, $2z$ or no folding)}
\label{fig:compmemstencil}
\end{figure*}

\begin{figure}
  \includegraphics[width=\columnwidth]{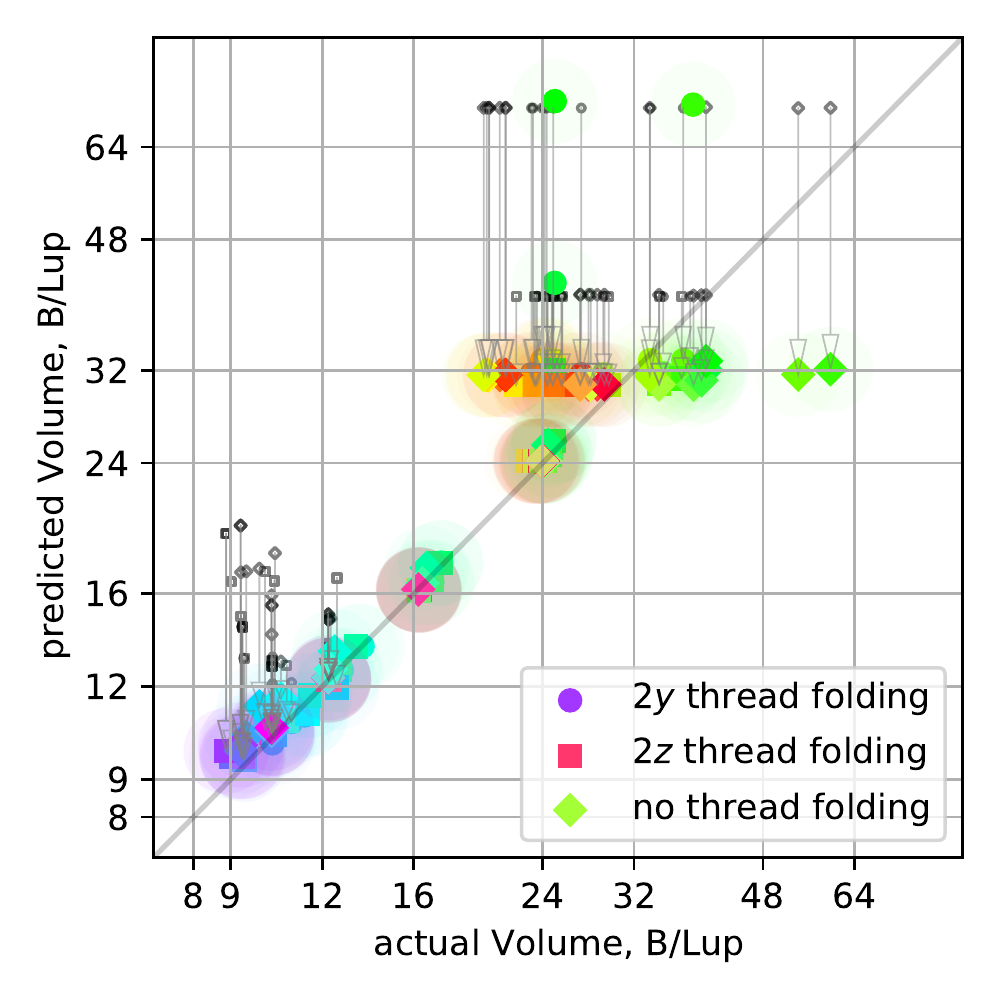}
\caption{Prediction vs measurement of DRAM load data volumes for the long range star stencil. Gray comparison markers show the change when considering hits in overlapping volumes.}
  \label{fig:memstencil}
\end{figure}

\begin{figure}
  \includegraphics[width=\columnwidth]{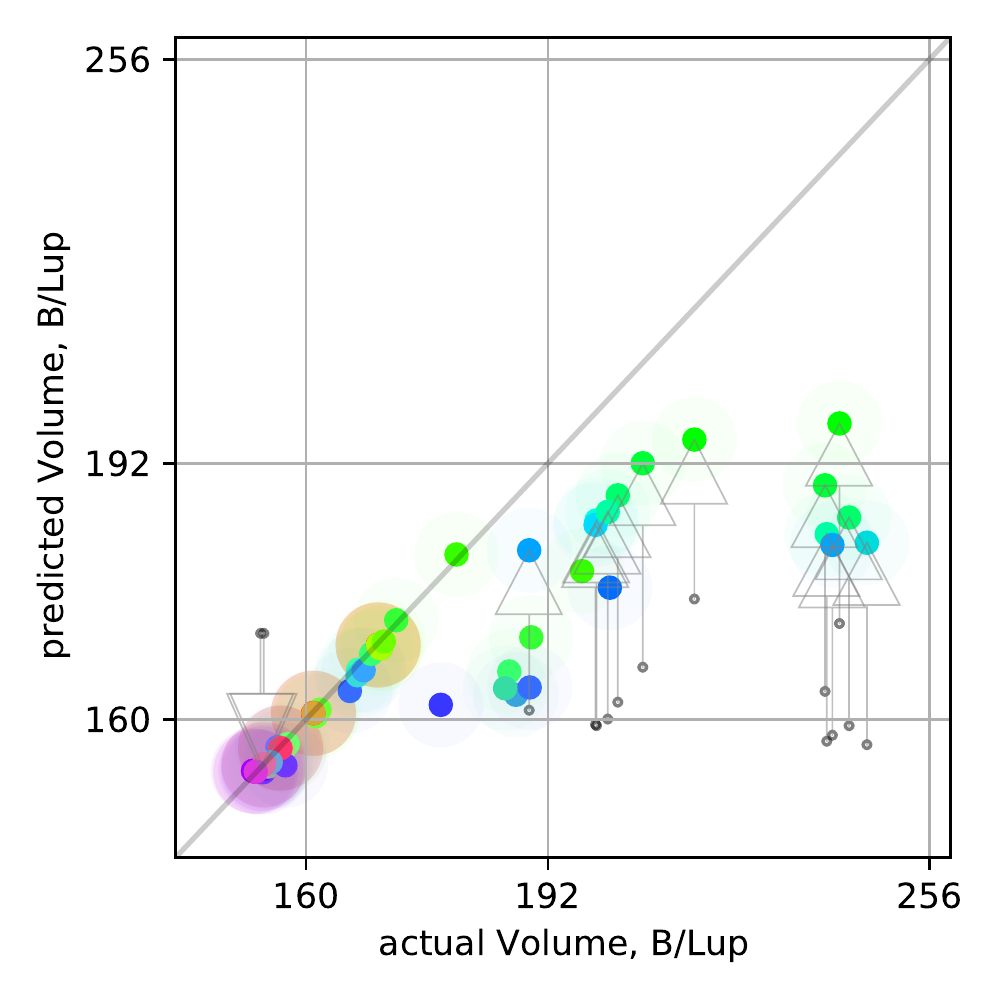}
\caption{ Prediction vs measurement of DRAM load data volumes for the LBM kernel. Gray comparison markers show the change when including capacity store misses.}
  \label{fig:memlbm}
\end{figure}

\begin{figure*}
  \centering
  \includegraphics[width=\textwidth]{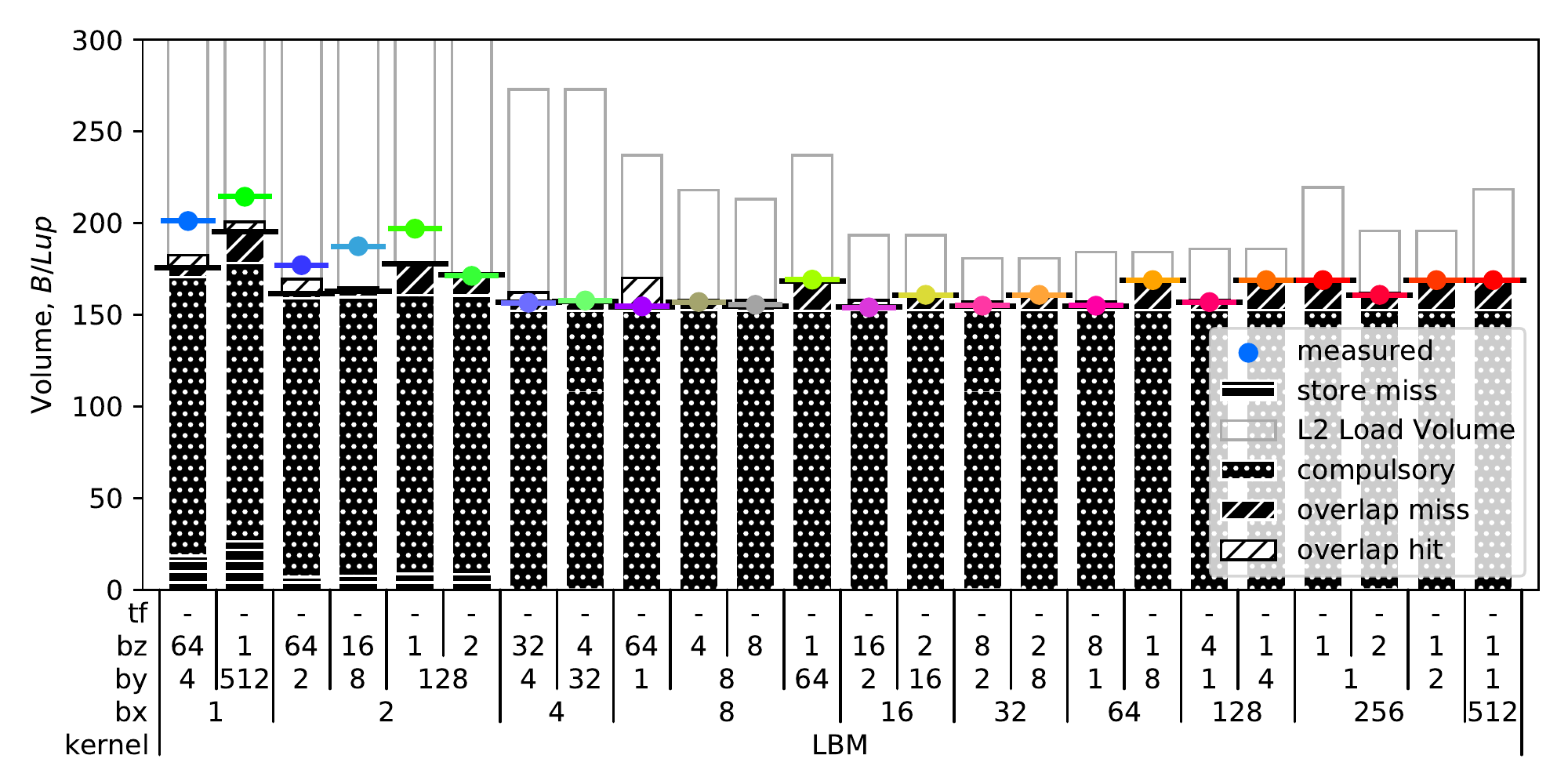}
\caption{Composition of stencil DRAM load data volumes for selected block sizes $(bx, by, bz)$. }
\label{fig:compmemlbm}
\end{figure*}

Figure \ref{fig:memstencil} shows a comparison of the memory load volume predictions of all stencil data points against the measured volumes.
The plot shows that thread blocks can be sorted by color into three categories, which are also represented in the detailed breakdown by component in figure \ref{fig:compmemstencil}.
Purple or violet colors, i.e. wide and deep thread block sizes with large $x$ and $z$ dimensions, e.g. $(32,1,32)$, have the lowest volumes.
Yellow/orange colors, representing wide and tall thread block sizes with large $x$ and $y$ dimensions, e.g. $(32,32,1)$, have larger data volumes.
Green, tall thread blocks with a large $y$ component, e.g. $(2,512,1)$, have the highest data volume.

The deeper thread block sizes, i.e. large z dimension, result in a lower balance because the grid is filled with thread blocks in $x,y,z$ order.
Most of the time a wave consists only of one layer of thread blocks in the $z$ direction, making the $z$ extent of the whole wave entirely dependent on the depth of a single thread block.
A very shallow wave, i.e. low $z$ extent,  results in little reuse in $z$ direction and hence high volumes.

The effects of considering hitting in the y and z layer overlapping volume is depicted with the gray markers and arrows.
The deep thread block sizes in purple and turquoise colors, which result in wave shapes comprising multiple z layers, profit most from the y layer reuse.
These shapes already have high reuse in the z dimension and the data volume of their z layer thread set is too large to still hit.
The lowest memory volumes of around $9 B/Lup$ are already close to the minimum of $8B/Lup$.

The group of shallow thread block sizes in red, yellow and green, profit a lot from z layer overlapping.
The estimate puts a lot of the thread block sizes on the same level, although the measurement shows a wider spread.
This is the same effect that was visible in figure \ref{fig:roverz}, where the red and yellow thread block sizes over perform and the green ones under perform the fitting curve.

Thread folding in the right dimension can reduce the compulsory data volume, see e.g. $(256,1,4)$.
A $2y$ thread folding is counterproductive here, as the $y$ component of the wave was already large.
The resulting large overlap with the previous wave does not come to fruition due to the increased L2 cache allocation and correspondingly increased capacity miss rates.
The $2z$ thread folding instead doubles the $z$ extent of the wave from four to eight, which improves the surface volume ratio of the wave and leads to a smaller compulsory volume.

Figures \ref{fig:memlbm} and \ref{fig:compmemlbm}   show the same data for the LBM kernel.
The streaming nature of the LBM part of the kernel leads to a much smaller dependence on the shape of the thread block size than the stencil.
There is a large cluster of shapes at around $150 B/Lup$, of which $120 B/Lup$ are due to the D3Q15 LBM kernel.

Instead, the memory data volumes for the LBM kernel are more influenced by the thread block x dimension.
The shorter the $x$ extent of a thread block, the fewer complete cache lines are loaded by a thread block in comparison to the partial cache lines loaded at the thread block boundary due to the unaligned LBM component loads, and the larger the amount of redundant loads.
This is even more valid for $x=1,2$, where all loads access only partial cache lines.
For these thread block sizes, there is also a large portion of partial stores that miss in the L2 cache.
Because the partially written cache line have to be read from the DRAM to complete the unwritten part of the cache lines, this partial store misses appear in the loaded data volume.

In the component breakdown, it is also clearly visible the there is a large compulsory data volume due to the streaming, and only comparatively small reuse opportunities.
The overlapping thread set volumes are usually to large to still hit, like for example the thread block size $(512,1,1)$.
There would be some overlapping hit potential, but the hit rate is zero.
On the other hand, a thread block size like $(8,1,64)$ successfully hits in the y layer overlap.

Shapes like $(64,1,8)$ have the lowest DRAM data volume due to their favorable wave shape, and not because of reuse from past waves.

\subsection{Layer Condition Like Phenomenon on GPUs}

\begin{figure*}
  \centering
  \includegraphics[width=\textwidth]{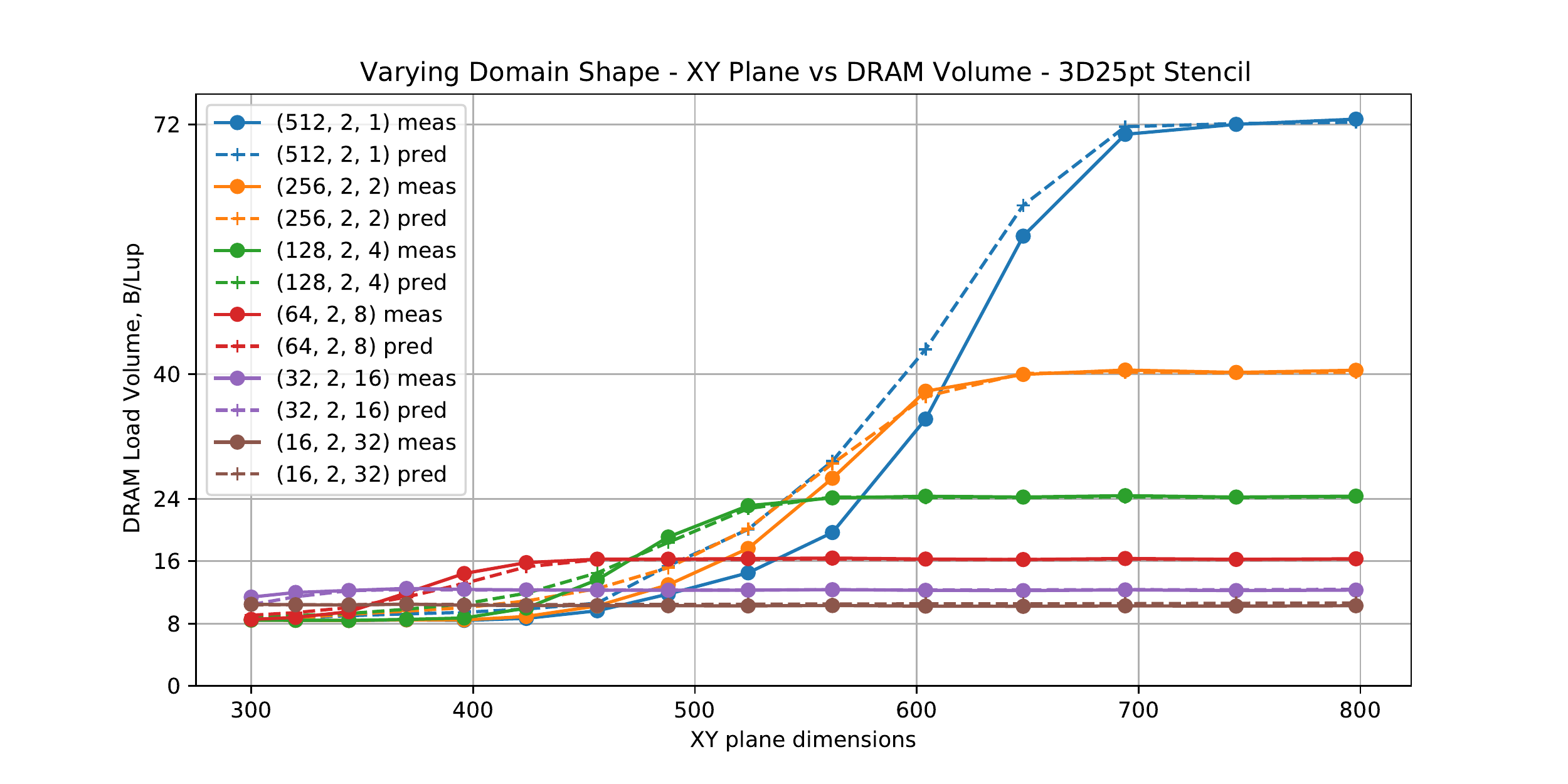}
  \caption{DRAM data load data volumes and predictions for the 3D25pt stencil for a series of domain sizes with quadratic, increasing $XY$ plane sizes. The depth of the domain is chosen as $z = 512 \cdot 1024 \cdot 1024 / (x \cdot y)$, so that the total size stays constant. }
  \label{fig:sizescan}
\end{figure*}

Figure \ref{fig:sizescan} shows a measurement series of the DRAM data volume for different domain sizes.
The domain sizes all have the same total size, but a different, quadratic $XY$ plane.

The different thread block sizes are chosen to have increasing $z$ extent.
For all the thread block sizes, there is a clear transition with increasing $XY$ extent from a small data volume almost as low as the absolute minimum of $8B/Lup$, up to a higher level that differs for each thread block size.

This is the transition from high to no reuse in the $z$ layer overlapping thread set due to growing $z$ layer data volumes.
With no $z$ layer reuse, all reuse has to come from the wave shape itself.
With 1/2/4/8/16/32 layers deep wave shapes, $\frac{9}{1}$/$\frac{10}{2}$/$\frac{12}{4}$/$\frac{16}{8}$/$\frac{40}{32}$ values have to be loaded from the DRAM, resulting in a volume of $72$/$40$/$24$/$16$/$10$ $B/Lup$.

The sequential layer condition is fulfilled for $X,Y <\sqrt {10\,MB / (9\cdot 8B)} = 381$, and the $(512,2,1)$ curve starts to increase just above 400.
The deeper $z$ extents of the wave shapes of the other thread block sizes increase the volume of the z layer thread set, or, formulated in the framework of the layer condition, require more than just the 9 layers to fit into the cache.
Correspondingly, the deeper thread block sizes run out of cache capacity earlier.

The predictions of the data volumes track the measured data volumes closely through the transition.
The domain size of $640\times512\times512$, which was used for the other measurements in this paper, would correspond to a quadratic $XY$ plane between $500^2$ and $600^2$.
The graph shows that this is a challenging domain size, as different thread block sizes are in different stages of their transition.

For performance prediction, this also means that the perfect configuration depends on the domain size.
For domains with small frontal aspect, the shallow thread block sizes have the lowest volume, whereas for larger domain sizes, the deeper thread block sizes are better.

\subsection{Performance Prediction}

\begin{figure} [h]
  \centering
  \includegraphics[width=\columnwidth]{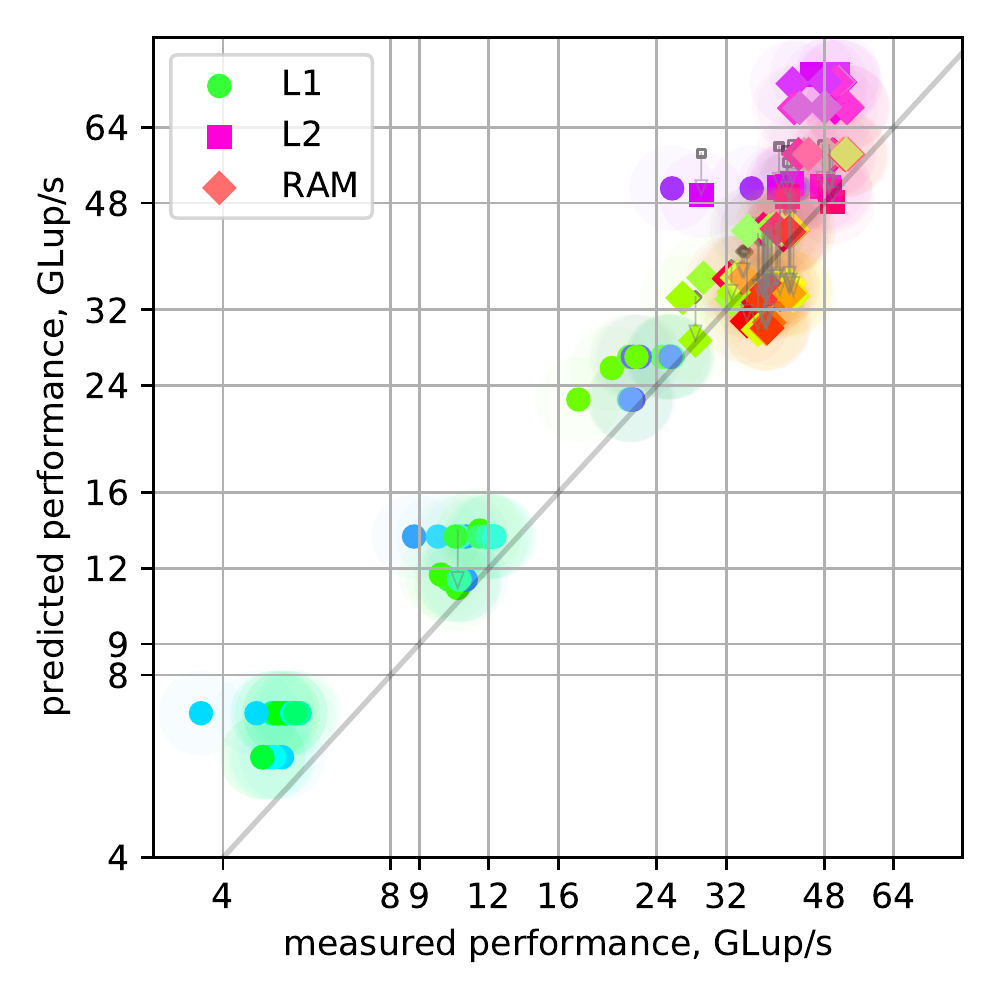}
  \caption{Performance prediction vs.\ measurement. The gray comparison
    markers and arrows show the difference to a phenomenological
    prediction made using the same performance model but with measured data volumes. 3D25pt/range 4 star stencil}
  \label{fig:rooflinestencil}
\end{figure}

\begin{figure} [h]
  \centering
  \includegraphics[width=\columnwidth]{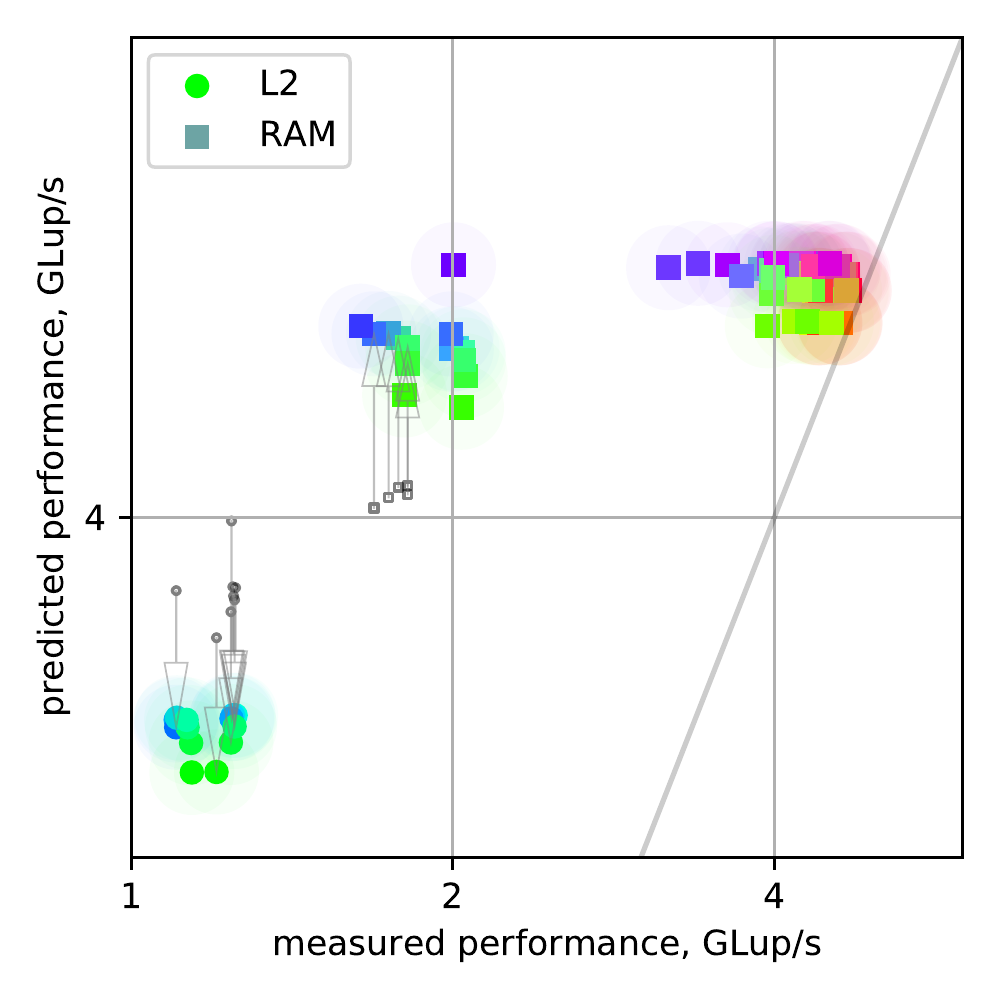}
  \caption{Performance prediction vs.\ measurement. The gray comparison
    markers and arrows show the difference to a phenomenological
    prediction made using using the same performance model but with measured data volumes. LBM kernel}
  \label{fig:rooflinelbm}
\end{figure}

Figures \ref{fig:rooflinestencil} and \ref{fig:rooflinelbm} show
comparisons of the predicted performance using the estimated data
volumes as input for the presented performance model.  The gray
comparison markers show the difference to a phenomonological
prediction that uses the same performance model but measured
data volumes.
For both applications, the performance model shows
overprediction, regardless of whether estimated or measured
data volumes are used.

For the 3D25pt range-four star stencil, the predictions are able to
rank the different configurations by performance and clearly captures
the performance differences between well-performing and badly-performing
configurations.
The thread block size $(16, 2, 32)$ with $2z$ thread folding is predicted
to be the fastest configuration, whereas measurements show it to be the 12th fastest.
However, it performs at $96\%$ ($50.7 GLup/s$) of the fastest measured configuration, which
is $(64, 4, 4)$ with $2z$ thread folding at $53 GLup/s$.

The original version of this paper (\cite{SBAC}) found the thread block size $(32, 2, 16)$ to be the optimal thread block size on the V100.
The larger L2 cache of the A100 shifts the transition points shown in the previous section, so that less wave shape inherent reuse is required.

The inability to find the actual fastest configuration is not due to
inaccurate data volumes, as a phenomenological performance model
using measured data volumes picks the same configuration as the
fastest.  The performance model cannot resolve 
the differences among the collection of well-performing configurations.
It does, however, correctly identify the
general type of configuration that performs well, as the best-predicted
and best-measured configuration do have similar shapes.

This is a relevant accomplishment, which is illustrated by comparison with
the thread block sizes found in \cite{higherorderstencils}. They find that among the
sizes $(8,8,8)$, $(16,16,4)$ and $(32,32,1)$, the
first one performs best for the application of a very similar
stencil to the one we used.
For measurements on the V100, that was used in that paper, we find
the best predicted thread block size $(16,2,32)$ to be
$36\%$ faster than a $(8,8,8)$ thread block size.  It is a nonintuitive
insight that a $(16,2,32)$ thread block size performs better
than the $(8,8,8)$, $(16,16,4)$ or $(32,32,1)$ thread block sizes they
have selected.

For this stencil, the most important limiter, especially with the
fastest configurations, is the DRAM bandwidth.  Although fewer
configurations are limited by the L2 cache bandwidth, the thread block
sizes with the lowest DRAM balance like $(32,1,32)$ are L2 cache
bandwidth limited because of their flat shape.  The L1 cache
limitation comes into play for thread block sizes with short
$x$ dimensions.

For the LBM kernel, the performance model only manages to correctly
identify the worst-performing configurations with short $x$
dimensions.  Apart from that, it cannot distinguish the averagely
performing from the well performing configurations.  Just as with the
stencil kernel, the fault is not with the estimated data volumes, but
with the performance model, which does not capture the relevant
mechanisms here. There is only a small group of thread block sizes with short
$x$ extent, that is limited by the L2 cache bandwidth.
For all other thread block sizes,
the LBM kernel, due to its streaming nature, is
limited entirely by the DRAM bandwidth.

\section{Conclusion and Outlook}

We have demonstrated an automated performance modeling process
for loop kernels on GPUs that is based on tracking data
accesses via address expressions. 
Our method can
estimate the data volumes transferred between the levels
of the memory hierarchy with high accuracy.  Its versatility
for a wide range of GPU programs has been demonstrated by
evaluating it with two challenging and diverse kernel types, a
long-range star stencil and a complex LBM kernel with a mix of access
characteristics.

We have shown the usefulness of these data volumes to gain insight
into the performance characteristics of a program and to classify
different code generation configuration by their performance.
However, our evaluation also showed that the simple performance model
does not capture all the performance relevant mechanics and fails to
differentiate between configurations at the top of the ranking.
Identifying and modeling these mechanisms is an important topic for
future work.
Modeling Translation Lookaside Buffer (TLB) misses would be one of the
candidates, where the relevant program metric would be TLB pages
accessed by the current wave.

We are creating a performance model that moves beyond the simple lowest limiter style model of the roofline model, but instead incorporates latency effects.

Another topic of future work is the testing and extension for
different hardware architectures.  The general hardware model with a
local L1 cache and a shared L2 cache is applicable for all current GPU
architectures, but details like cache line sizes and cache
capacities have to be adapted.  For example, AMD's current CDNA
architecture's much smaller L1 cache would lead to many more capacity
misses.

In the future, we export more complex cache hierarchies, with more cache levels that are distributed over multiple partitions and chips.
In this work, we have simply assumed full cache line duplication and halved capacity to account for the split L2 cache on the A100 GPU.
It is possible to instead estimate the amount of cache line duplication and therefore the amount of effective cache capacity.
Such an estimate would also be closely related to an estimate for the amount of cache traffic between the cache sections, which we have disregarded as a performance limiter in this work.

We are also looking to verify the applicability of our
method for more applications and more complex code transformations
like temporal blocking.
We want to investigate advanced grid iteration schemes like spatial blocking, which is a typical stencil optimization on CPUs.
Our results have shown the CPU like layer condition behavior, which also shows that spatial blocking is applicable.
Spatial blocking effectively reduces the frontal aspect of the domain, to fulfill the layer condition.
For temporal blocking and other complex stencil iteration schemes,
our solution could be used to choose parameters like blocking factors and parallelization schemes.

Similarly to the Layer Condition Calculator website (\cite{layerconditionexplorer}), we plan to crate a simple web interface that allows to type in address expressions by hand and computes estimates for the given parameters.

The authors gratefully acknowledge the scientific support and HPC resources provided by the Erlangen National High Performance Computing Center (NHR@FAU) of the Friedrich-Alexander-Universität Erlangen-Nürnberg (FAU). The hardware is  funded by the German Research Foundation (DFG).

The code of the \emph{Warpspeed} data volume and performance estimator is available at \cite{warpspeed}.

\bibliographystyle{IEEEtran}
\bibliography{references}

\end{document}